# Angular Dependence of Hump-shape Hall Effects for Distinguishing between Karplus-Luttinger and Geometrical Origins

*Zhi Shiuh Lim*[1,#], *Lin Er Chow*[1], *Amy Khoong Hong Khoo*[2], *Ganesh Ji Omar*[1], *Zhaoyang Luo*[1], *Zhaoting Zhang*[1], *Hong Yan*[1], *Ping Yang*[3], *Robert Laskowski*[2], *A. Ariando*[1,*]

1. Physics Department, Block S12, #2 Science Drive 3, National University of Singapore 117551

2. Institute of High-Performance Computing (IHPC), A*STAR, Fusionopolis Way, #16-16 Connexis, Singapore 138632

3. Singapore Synchrotron Light Source (SSLS), National University of Singapore, 5 Research Link, Singapore 117603

Email: *ariando@nus.edu.sg, #mpev527@visitor.nus.edu.sg

**Abstract:**

**Among the vast magnetic heterostructures explored in Condensed Matter Physics, two contrasting interpretations of the hump-shaped Hall Effects remain ambiguous and debated, namely, the overlap of two opposite-signed Karplus-Luttinger Hall loops associated with inhomogeneous collinear domains with perpendicular anisotropy, or the Geometrical/Topological Hall Effect emanated from hexagnal close-packed lattice of Skyrmion ground state with smoothly varying non-collinear moments. Their similarity in topology implies difficulty in discrimination via magnetic imaging. Here, this ambiguity is overcome and clarified by the divergence exponent of hump peak fields extracted from**

**Hall measurements with magnetic field rotation on several heterostructures. Their difference in sensitivity to in-plane fields reveals that the former mechanism involves higher uniaxial anisotropy than the latter, departing from the Skyrmion ground state regime by the Ginzburg-Landau framework of triple-*q* spin-wave superposition. Numerous material systems can be summarized into a single curve of divergence exponent versus the collinear quality factor, bridging the crossover of the two mentioned mechanisms.**

**Introduction:**

Theoretical development on Anomalous Hall Effect (AHE) began in 1954 when Karplus and Luttinger (KL)[1] formulated the perpendicular electron velocity of the intrinsic mechanism in simple ferromagnetic metals, $v_{\perp} = \frac{1}{i\hbar}\left[\hat{\mathrm{H}}_{\mathrm{soc}}, x_{\perp}\right] = \frac{1}{m^2c^2}\left(\frac{\boldsymbol{M}}{M_s} \times \boldsymbol{\nabla} V\right)$, where spin-orbit coupling (SOC) is the crucial perturbation to the system subjected to an external electric field ($\nabla V$). Later refinement of KL-AHE by McDonald and Nagaosa et al. incorporated the language of k-space Berry curvature $\boldsymbol{\Omega}(\boldsymbol{k})$ to account for anti-crossing in band structures[2]. Whereas the extrinsic mechanisms involving impurity scattering were account by classifications into skew scattering[3] and side-jump[4]. Until the last two decades, these three sideway electron deflection mechanisms, all of which require SOC on a uniform collinear magnetization ($\boldsymbol{M}$) background, formed the backbone of conventional AHE analyses via $\rho_{xy}^{\mathrm{AHE}}(H) \propto {\rho_{xx}}^{\beta} M(H)$ where $1 \leq \beta \leq 2$ is the extracted exponents[5]. Then, the abrupt enhancement of non-linear $\rho_{xy}(H, T)$ near Curie temperature ($T_C$) observed in perovskite manganites[6] contrasted with the conventional AHE mechanisms which is expected to diminish with increasing temperature. This motivated extensive theoretical study into the unconventional *subset* of AHE – Geometrical Hall Effect (GHE) emanating from free electrons interacting with chiral magnetic moment textures with a net solid angle $\Omega = 2\tan^{-1}\left[\frac{\widehat{\boldsymbol{m}}_i \cdot (\widehat{\boldsymbol{m}}_j \times \widehat{\boldsymbol{m}}_k)}{1 + \widehat{\boldsymbol{m}}_i \cdot \widehat{\boldsymbol{m}}_j + \widehat{\boldsymbol{m}}_j \cdot \widehat{\boldsymbol{m}}_k + \widehat{\boldsymbol{m}}_k \cdot \widehat{\boldsymbol{m}}_i}\right]$. In the regime of strong and adiabatic exchange interaction, Ω defines the real-space Berry phase gained by the mobile electrons, which evolves into the Berry curvature or pseudo magnetic field $B_{\mathrm{eff}} = \frac{1}{4\pi}\int \boldsymbol{m}\cdot\left(\frac{\partial \boldsymbol{m}}{\partial \mathrm{x}}\times\frac{\partial \boldsymbol{m}}{\partial \mathrm{y}}\right)\mathrm{dxdy}$, resulting $\rho_{xy}^{\mathrm{GHE}}(H) \propto \frac{B_{\mathrm{eff}}}{ne}$ in analogy to the ordinary Lorentz force effect (OHE)[7]. Theoretically, GHE does not require SOC[8] in contrast to the conventional AHE, although the Skyrmion formation may require SOC-induced Dzyaloshinskii-Moriya Interaction (DMI)[9] or crystallographic frustration[10]. Yet the presence/absence of SOC in electron deflection is difficult to verify in experiments.

Any magnetic texture can be described in cylindrical coordinates ($r$,$\phi$) as:

$$\boldsymbol{m}(r,\phi) = \begin{bmatrix} \sin(f(r))\cos(Q_v\phi + Q_h) \\ \sin(f(r))\sin(Q_v\phi + Q_h) \\ \cos(f(r)) \end{bmatrix} \quad [1]$$

where $Q_{v,h}$ are the vorticity and helicity[11]. Notably, the radial profile $f(r)$ that varies within the range of {0,$\pi$} exhibits a crossover between skyrmion ground state and collinear domains bordered by chiral domain walls. On one hand, a smoothly varying sinusoidal $f(r)$ typically describes a hexagonal close-packed Skyrmion-lattice (SkL) with completely non-collinear moments stabilized at nonzero out-of-plane magnetic fields, hence the existence of $B_{\mathrm{eff}}$ is valid. Besides, in the atomic-scale SkL regime[12], if the mobile electrons' mean free path ($\lambda_{\mathrm{mf}}$) is comparable or longer than the magnetic unit cell, the electronic $k$-space band structure would be modified into a massive Dirac Fermion[13], in analogy to the Haldane's model[14]. This is the more complete picture of GHE accountable by the Kubo formula in $k$-space although collinear moments are completely absent. Yet if electrons hop across SkL with short $\lambda_{\mathrm{mf}}$, the massive Dirac Fermion band structure should vanish, the GHE theory should follow the Boltzmann transport across non-coplanar spin-clusters[8b]. On the other hand, collinear domains are described by a more step-function-like $f(r)$, i.e.: the area ratio of collinear moments to chiral domain walls (DW) is large. Then, large $M_{\mathrm{sat}}$ materials may stabilize isolated bubble skyrmions dispersed in a ferromagnetic matrix or bubble-lattice with intermediate squareness of $f(r)$ at zero magnetic field. These entities are stabilized by dipolar/demagnetization energy and found at non-ground states (metastable valleys) of free energy[15]. Hence, a large KL-AHE contribution can be expected in these bubbles since regions of collinear magnetization are significant. Conversely, the $\left(\frac{\partial \boldsymbol{m}}{\partial \mathrm{x}} \times \frac{\partial \boldsymbol{m}}{\partial \mathrm{y}}\right)$ term in $B_{\mathrm{eff}}$ should diminish since the surrounding chiral DWs would be nearly straight and too long compared to the electrons' mean-free path, leading to difficulty in collecting the real-space Berry phase[16].

As mentioned, in the effort of distinguishing Hall-humps due to GHE or overlapping KL-AHE, ambiguities and debates arise from at least two factors: (1) both the SkL ground state and metastable Skyrmions can be described by eqn. [1] thus having similar topology; (2) GHE and KL-AHE only differ subtly by their requirement of SOC in deflecting mobile electrons. GHE is usually recognized by its hump-shape $\rho_{xy}^{T}(H_z) \sim \mathcal{L}(H_{\mathrm{z}})/H_{\mathrm{z}}$ where $\mathcal{L}(H_{\mathrm{z}}) = \coth(H_{\mathrm{z}}) - 1/H_{\mathrm{z}}$ is the Langevin function, and has been observed in various heterostructures of oxide[17] or magnetically-doped topological insulator (MTI)[18] bilayers. However, the Hall-humps can also be fitted by overlapping two KL-AHE loops with opposite signs, implying that the material systems in question contain inhomogeneous collinear domains with opposite *k*-space Chern numbers, doubting the existence of skyrmions. Intensive efforts were invested on magnetic imaging to investigate these low-temperature and low-magnetization materials using scanning probe techniques[19] and X-ray Photoemission Microscopy (XMCD-PEEM)[20]. Notably, in single-layer ultrathin $SrRuO_3$ (SRO) with Hall-humps, terrace engineering was used to segregate/coalesce the bubbles along terraces[21], to be in favour of the KL-AHE interpretation. On the other hand, the $SrRuO_3/PbTiO_3$ bilayer was found to host a double-$\boldsymbol{q}$ square meron lattice[22] which supports the GHE interpretation. This is similar to the MTI bilayers which have been understood to stabilize SkL via interfacial antiferromagnetic exchange[23], and antiferromagnetic frustration in $Gd_2PdSi_3$[10b]. Hence, a stereotypical viewpoint on Hall-humps should be avoided; yet further angle-dependence and domain wall energy analyses on the Hall-humps in oxide films were lacking. In this work, by applying magnetic field rotation on contrasting samples, we show that collinear domains and ground state SkL can be distinguished by relating the divergence of their Hall-hump peak fields ($H_{peak}$) to domain wall energies ($\sigma_{DW}$). The high sensitivity of SkL-hosting samples to in-plane magnetic field is understood from Ginzburg-Landau theory of triple-$\boldsymbol{q}$ spin-wave superposition and supported by micromagnetic simulation. The $H_{peak}$ divergence exponent ($\gamma$) from various heterostructures also forms a linear trend with the quality factor constituting a continuous crossover from SkL ground state to collinear domains. Such analyses may potentially become a useful protocol for future disambiguation.

**Results:**

Figures 1a,b and S1a,b present the selected oxide heterostructures for analyses, with fabrication details described in *methods*. Perovskite SRO thin films are known to have two distinct monoclinic (m-) and tetragonal (t-) phases, with Glazer notations[24] of octahedral rotations/tilts distinguished as $a^-b^+c^-$ and $a^0a^0c^0$ respectively from X-ray Bragg diffractions (XRD) around half-integer HKL-indices[25] (Figure S1c). Their KL-AHE loops also differ significantly, i.e. negative- and positive-sign for t-SRO and m-SRO respectively (Figure S1a,b). This difference originates from the opposite signs of *k*-space Chern numbers of different 4d topological $t_{2g}$-bands, related to the different band occupancy at Fermi level and lower saturation magnetization ($M_{sat}$) of t-SRO than m-SRO[17b, 21]. In Figure 1a, the "$mR_3tR_7$" structure formed by ultrathin m-SRO(3uc)/t-SRO(7uc) on $SrTiO_3$(001) exhibits obvious Hall-humps ranging from 3 K to 40 K, where "uc" refers to "pseudo-cubic perovskite unit-cell thickness". It assimilates a z-direction phase separation, intended to represent many recent publications around single-layer SRO including those with xy-plane inhomogeneity[17b, 21, 26]. This is generally valid since an ultrathin t-phase buffer layer is inevitable when SRO is interfaced directly with the cubic $SrTiO_3$(001) surface due to the suppression of octahedral rotations/tilts[27], regardless of the growth pressure or substrate's surface termination.

Meanwhile, the "$mR_5I_{10}$" structure in Figure 1b constructed by m-SRO(5uc) on $SrIrO_3$(10uc) also shows large Hall-humps across a wide temperature range of 10-90 K. Here the paramagnetic $SrIrO_3$ is believed to contribute strong SOC and DMI, while functioning as a buffer layer with thickness beyond the octahedral suppression region such that the m-SRO layer above can be proven to be free from phase separation issue. In Figure S2, we provide the magnetic force microscopy (MFM) images of $mR_5I_{10}$ at 20 K and varying out-of-plane magnetic field, proving that magnetic textures akin to SkL indeed exists and reaches maximum density at $H_{\text{peak}}$, agreeing well with the Hall-hump trend.

To shed light on the universality of our subsequent analyses, we included the 8uc and 3uc variants of $Tm_3Fe_5O_{12}$(TmIG) films capped with 2-nm Pt grown on $Gd_3Ga_5O_{12}$(111) substrates, labelled as "$P_2T_8$" and "$P_2T_3$", as displayed in Figure 2. TmIG(111) is renowned to be a high-$T_C$ ferrimagnetic insulator with perpendicular anisotropy (PMA) under tensile strain, hence the spin current reflected off the Pt/TmIG interface can detect clear Spin Hall Effect (SHE) induced square KL-AHE loops[28] in $P_2T_8$ and GHE[29] in $P_2T_3$ at 300 K. Being slightly different from $mR_3tR_7$ and $mR_5I_{10}$, the Hall-humps in $P_2T_3$ are non-hysteretic and are greatly enhanced around $T_C$~300 K of the ultrathin (3uc) TmIG, but vanishes at low temperatures. Such phenomenon has been discussed as the chiral fluctuation behaviour[17c] with a proximity-magnetized Pt in its carrier localization regime, and exhibits a power-law enhancement around the 2$^{\text{nd}}$-order transition temperature[30].

The main focus of this work is the investigation of Hall effects with rotation of magnetic field ($\rho_{xy}(H_{\text{total}}, \theta)$) from out-of-plane ($H\|z$ at θ=0°) approaching in-plane ($H\|x,y$ at θ→90°), at 20 K for $mR_3tR_7$ and $mR_5I_{10}$ and 300 K for $P_2T_8$ and $P_2T_3$. The measurement schematic is illustrated in Figure 1c, presented in total field, $H_{\text{total}} = \sqrt{{H_z}^2 + {H_{x,y}}^2}$, implying the field sweep is in the 1$^{\text{st}}$ and 3$^{\text{rd}}$ quadrants. Notably in Figure 1d, the $H_{\text{peak}}$ of $mR_3tR_7$ exhibits fast divergence with increasing θ (left panel), which can be fitted with a phenomenological $H_{\text{peak}} \propto$ $1/\cos^{\gamma}(\theta)$ to obtain γ=0.76 (Figure 1f). Note that the γ→1 limit implies a sample with high uniaxial magneto-crystalline anisotropy ($K_U$) that receives zero influence from in-plane fields since the out-of-plane magnetization component ($M_z$) is responsible for the KL-AHE. It is then viable to decompose the Hall data of $mR_3tR_7$ at any θ into two sigmoidal KL-AHE loops of opposite signs via $\rho_{xy}(H,\theta) = \sum_{i=1,2} A_i \left\{\coth\left[B_i\left(H \pm \frac{H_{C,i}}{\cos^{0.76}(\theta)}\right)\right] - \left[B_i\left(H \pm \frac{H_{C,i}}{\cos^{0.76}(\theta)}\right)\right]^{-1}\right\}$ where and $A_i$ and $B_i$ are coefficients, and the two coercive fields ($H_{C1,2}$) delimit the range where the Hall humps would emerge. The decomposition details are shown in *Supporting* Figure S6. Such θ-dependent Hall analysis, albeit unprecedented, supports that $mR_3tR_7$ hosts inhomogeneous collinear bubble domains at intermediate fields in agreement to reference [21]. Likewise, the KL-AHE of $P_2T_8$ in Figure 2a,c also showed an obvious divergence of $H_C \propto 1/\cos^{0.45}(\theta)$. In contrast, the $H_{peak}$ of $mR_5I_{10}$ and $P_2T_3$ were found to be independent of θ with extracted γ~0.06 and 0.03 while the hump magnitudes ($\Delta\rho_{xy}$) also diminished at large θ-angles, as shown in Figure 1e,f and Figure 2d-f respectively.

To understand the large divergence reflected by γ=0.74 and 0.45 in $mR_3tR_7$ and $P_2T_8$ respectively, we further resolve the out-of-plane ($H_z$) and in-plane ($H_{x,y}$) field components during measurements. The Hall measurements were done by sweeping $H_z$ with a constant $H_{x,y}$, while the $H_{total}$ vector is rotating in the 1st and 2nd quadrants or the 4th and 3rd quadrants for a particular $\rho_{xy}(H_z, H_{x,y})$ loop (schematic in Figure 3e). In Figure 3a,b, a linear shift of z-component peak field $H_{peak,z}$ towards left and right sides can be observed in $mR_3tR_7$ with varying $H_{x,y}$ in the range of ±3 T. Notably, the loop width demarcated by the $\pm H_{peak,z}$ stayed nearly constant, thus the $\mp H_{peak,z}$ opposite to the shift direction can reach near-zero at large $\pm H_{x,y}$. Similar shift in z-component ($H_{C,z}$) is evident in $P_2T_8$ as shown in *Supporting* Figure S3a,b and text. Such loop-shift can be understood as a destabilization of the $\pm M_z$ by the in-plane magnetic field, ie.: easier (harder) to switch $M_z$ downwards (upward) assisted by a positive $H_{x,y}$, and vice versa. Hence, the horizontal expansion of $\rho_{xy}(H_{total},\theta)$ and the horizontal shift of $\rho_{xy}(H_z, H_{x,y})$ observed in the two field-rotation schemes are equivalent, since the former involving field-sweep in the $1^{st}$-$3^{rd}$ (or $4^{th}$-$2^{nd}$) quadrants causes increasing switching difficulty to both polarity of $\pm M_z$. The data inter-conversion in these two schemes can be seen in *Supporting* Figure S4a,b. Note that a broken parity symmetry exists where a right-shift of $\rho_{xy}(H_z, H_{x,y})$ loop is always observed for $+H_{x,y}$ and vice versa. This can be understood by the large polar angle of SRO's easy axis tilting up to 45° away from [001] due to a competition between in-plane and out-of-plane anisotropies[25, 31] (Figure 3c); and is also similar to $P_2T_8$ (Figure S3a). Such broken parity is confirmed to be irrelevant to the spin-orbit torque (SOT), since neither of loop shift direction reversal upon $J_x \Leftrightarrow J_{-x}$ nor loop shift vanish with changing $H_x \Rightarrow H_y$ was observed. To be consistent with $mR_5I_{10}$, we limited our investigation to magnetostatic at low current densities to avoid magnetization dynamics and Joule heating; however, the loop shift enhancement can be expected if larger $J_x$ up to $10^{10}$ A/m$^2$ is sourced, as exemplified by $P_2T_8$ (Figure S4c). Conversely, under the same "resolved field-components" scheme, $mR_5I_{10}$ (Figure 3d) and $P_2T_3$ (Figure S3c) showed stationary $H_{z,\mathrm{peak}}$ without left/right shifting but fast-vanishing Hall-humps with varying $H_{x,y}$. Hence, it is convincing to rule out the existence of inhomogeneous collinear bubbles in $mR_5I_{10}$ and $P_2T_3$.

For comparison with properties of Néel-type SkL, we first evaluated the interfacial DMI values of $mR_3tR_7$ and $mR_5I_{10}$ via Density Functional Theory (DFT) calculations, by using the "two-slab" structural models as displayed in Figure 4a,b. The mismatches in octahedral tilt across the interfaces are accounted, i.e. SRO($a^-b^+c^0$)/SRO($a^0b^0c^0$) for $mR_3tR_7$ and SRO($a^-b^+c^0$)/SIO($a^+b^0c^0$) for $mR_5I_{10}$, as deduced from half-integer XRD data (Figure S1c). The obtained diffraction peaks agree qualitatively with the expected concepts of bond length variation[32] and strain relaxation as discussed in *Supporting Information*, although the quantitative rotation angles with thickness variation cannot be determined at present. As shown in Figure 4c,d, due to the presence of heavy element ($Ir^{4+}$), a significantly higher interfacial

DMI of 0.76 mJ/m$^2$ was obtained in $mR_5I_{10}$ as compared to merely 0.071 mJ/m$^2$ in $mR_3tR_7$, contrasting their likelihood in hosting skyrmions. Hence, MUMAX$^3$ micromagnetic simulations[33] were performed on $mR_5I_{10}$ to compute the topological charge density (*TCD*) = $\frac{1}{4\pi}\boldsymbol{m}\cdot\left(\frac{\partial\boldsymbol{m}}{\partial x}\times\frac{\partial\boldsymbol{m}}{\partial y}\right)$, by using the DMI and other parameters extracted from magnetometry (Figure S5b) as inputs, and incorporating magnetic field rotation $B_{\text{ext}} = B_{\text{total}}[\sin\theta, 0, \cos\theta]$ (*Supporting Information*). In Figures 4e, the $TCD(H_{\text{total}}, \theta)$ mappings shows that the peak fields corresponding to the densest SkL remain non-diverging, in good agreement to the $\rho_{xy}(H_{\text{total}}, \theta)$ trend shown in Figure 1e. The similar observation was obtained in $P_2T_3$ by using parameters extracted from various references (Figure S3d). This can be understood by considering the Ginzburg-Landau framework for triple-$\boldsymbol{q}$ spin-waves superposition. It is well-known that a trio of cycloidal spin-waves that can be described by $\boldsymbol{m}_i = \sum_{i=1,2,3}[\hat{\boldsymbol{z}}\cos(\boldsymbol{q}_i\cdot\boldsymbol{r}) \pm \boldsymbol{q}_i\sin(\boldsymbol{q}_i\cdot\boldsymbol{r})]$ would create a two-dimensional (2D) hexagonal close-packed Néel-SkL stabilized by a small $H_z$, where $\boldsymbol{r} = [x, y, z]$, and wave vectors $\boldsymbol{q}_i = \left[\cos\left(\frac{2\pi i}{3}\right), \sin\left(\frac{2\pi i}{3}\right), 0\right]$. Likewise, helicoidal spin-waves with $\boldsymbol{m}_i = \sum_{i=1,2,3}[\hat{\boldsymbol{z}}\cos(\boldsymbol{q}_i\cdot\boldsymbol{r}) \pm (\hat{\boldsymbol{z}}\times\boldsymbol{q}_i)\sin(\boldsymbol{q}_i\cdot\boldsymbol{r})]$ would create a 2D Bloch-type SkL. Then, application of $H_x$ would cause $\boldsymbol{m}$ to precess around $H_x$, hence the dominant single-$\boldsymbol{q}$ is perpendicular (parallel) to $\boldsymbol{H}_x$ for the Néel-type (Bloch-type) case[34]. Due to this vulnerability of SkLs subjected to in-plane fields, the total field of peak *TCD* should be almost θ-independent before skyrmion annihilation. Figure 5a illustrates that the variation of $K_U$ within one order-of-magnitude in the simulation creates gentle *TCD* peak-field divergence around γ~0, hence is unable to produce large divergence like that of $mR_3tR_7$. These simulations do not lose generality and are invariant under the change of $H_x \Leftrightarrow H_y$, while both Néel-SkL and Bloch-SkL show the same vulnerability to in-plane fields in such thin film regime. The observations above validate the grouping of Hall-humps of $mR_3tR_7$ and $P_2T_8$ as indication of collinear bubble domains, while $mR_5I_{10}$ and $P_2T_3$ can be categorized as hosting skyrmions.

**Discussion:**

From the gigantic GHE observed in $mR_5I_{10}$ and $P_2T_3$, we could speculate that the SkL involves nanometer-scale magnetic unit cell, where each unit cell contains a unified topological charge without mutual cancellation, so that the emergent flux per unit area can be huge. SkL ground state is defined to contain purely non-collinear moments homogeneously distributed in two-dimension across the whole sample with negligible collinear region. If neglecting the dipolar energy, the SkL ground state is achieved when $\sigma_{\mathrm{DW}} = 4\sqrt{A_{\mathrm{ex}}K} - \pi|D|$ becomes negative[35], where $A_{\mathrm{ex}}$ is the exchange stiffness and $K$=$K_{\mathrm{U}}$. Whereas the second term ($\pi D$), which is crucial for lowering $\sigma_{\mathrm{DW}}$, may originate from the SOC-induced DMIs or effectively from geometrical or antiferromagnetic frustration. Thus far, the two types of most well-known DMI are the helicoidal-DMI with energy density $-D_{hel}[\boldsymbol{m} \cdot (\boldsymbol{\nabla} \times \boldsymbol{m})]$ and the cycloidal-DMI with energy density $D_{cyc} m_z (\boldsymbol{m} \cdot \boldsymbol{\nabla} - \boldsymbol{\nabla} \cdot \boldsymbol{m}) m_z$. In materials with crystal structures with $D_{\mathrm{n}}$ symmetry such as the B20 compounds, $D_{hel}$ dominates at phase spaces of weak ferromagnetic properties, and $\pm D_{hel}$ stabilizes Bloch-type SkL with uniform (single-valued) vorticity ($Q_v$=1) and helicity $Q_h$=$\pm\pi/2$ at low magnetic fields. Likewise, $\pm D_{cyc}$ dominates in materials with $C_{n\mathrm{v}}$ symmetry such as the rhombohedral $GaV_4S_8$[36] and tetragonal $VOSe_2O_5$[37] and stabilizes Néel-type SkL (uniform $Q_v$=1, $Q_h$=0 or $\pi$). In such DMI-dominated regime (*Regime-4* defined here), labyrinthine stripes (spin-waves) must form at zero field instead of SkL. A small external magnetic field must be applied to allow superposition of stripes into SkL, as described by Ginzburg-Landau framework.

On the other hand, if dipolar energy is important, it is well-known that two length scales of collinear domains should be compared, i.e.: domain wall width $w = \sqrt{\frac{A_{\mathrm{ex}}}{K}}$, and domain size $l_{\mathrm{d}} = \frac{2\sqrt{A_{\mathrm{ex}}K}}{\mu_0 M_{\mathrm{sat}}^2} = \frac{\sigma_{\mathrm{DW}}}{4K_{\mathrm{shape}}}$ which is the ratio between $\sigma_{\mathrm{DW}}$ to dipolar energy. The quality factor of collinearity is $Q_c = \frac{l_{\mathrm{d}}}{w} = \frac{2K}{\mu_0 M_{\mathrm{sat}}^2}$[38]. In general, the $K$ term should be modified into: $K =$

$\begin{cases} K_\mathrm{U} - \mu_\mathrm{o} M_\mathrm{sat}^2/2 = K_\mathrm{eff} & , \quad \text{thin films} \\ K_\mathrm{U} + \mu_\mathrm{o} M_\mathrm{sat}^2 \cos^2(Q_h)/2 & , \quad \text{bulk samples} \end{cases}$, where the $\mu_\mathrm{o} M_\mathrm{sat}^2/2$ term accounts for the shape anisotropy ($K_{shape}$) with demagnetization tensors ±1, and $K_{eff}$ is the experimentally measured anisotropy from *M-H* curves (*Supporting* Table S2). Firstly, we may define *Regime-1* as the trivial large collinear domains bordered by thin Bloch-walls with high $Q_c$ and sharp step-function *f*(r). The modified *K* for bulk samples implies that dipolar energy increases the difficulty of entering the SkL ground state with the presence of a small field. Besides the Néel-type ($Q_h$=0,π) SkL ground state is more difficult compared to Bloch-type ($Q_h$=±π/2) SkL ground state since a higher $D_{cyc}$ compared to $D_{hel}$ is needed to achieve the equivalent effect. Then, for films with thickness $h < l_\mathrm{d}$, on top of reduced $K_{eff}$, choosing materials with large $M_{sat}$ would reduce $Q_c$, hence the area fraction of non-collinear moments increases. The textures qualify as "isolated Bloch-walled Skyrmion-like bubbles" that is metastable at zero-field dispersed in a ferromagnetic background (defined as *Regime-2*) albeit $D_{cyc}$ or $D_{hel}$ can be completely absent. Increasing $|D_{cyc}|$ would reduce $Q_c$ further, and a Néel-walled bubble lattice (*Regime-3*) would emerge at zero-field but still contains significant collinear moment regions. Finally, *Regime-4* with zero-field stripes and small-field SkL is reached for $\sigma_{DW} \leqslant 0$[15, 39]. In such phase evolutions, true GHE and TCD would become large only upon entering *Regime-4*; while the phase window of *Regime-2,3* would shrink to become unobvious in perovskite oxides due to intrinsically lower $M_{sat}$ from a larger atomic unit cell volume compared to real metals. Such evolution has not been well-clarified in many publications. Thus far, *Regime-2* is the more favourable phase than *Regime-3,4* in effort of developing the Skyrmion racetrack memory[40], and must be identified by a magnetic imaging technique. However, a trade-off exists since the significant collinear magnetizations may result in higher threshold current densities for movement driven by spin-torques[41].

Here, our γ-exponent scheme can be calibrated to quantify the mentioned phase evolution by a continuous trend. We repeated the total field rotation scheme for a plethora of other oxide heterostructures with data listed in *Supporting* Table S2. As summarized in Figure 5b, a linear scaling relationship between the γ-exponent and $Q_c$ was obtained, where $Q_c = \frac{2K}{\mu_0 M_{sat}^2} - \frac{\pi|D|}{2\mu_0 M_{sat}^2}\sqrt{\frac{K}{A_{ex}}}$ is a good representation of the sharpness of *f*(r). There could be difficulties in extracting the peak TCD field positions in micromagnetic simulations for Néel-SkL due to fast vanishing TCD with increasing θ, hence creating artefactual negative shifts in γ for the simulated results (red datapoints in Figure 5b). Nevertheless, the linear scaling is insightful. In particular, the $SrRuO_3/BaTiO_3$ bilayer and superlattice heterostructures can be found at the intermediate regime of the plot, while their magnetometry and γ extraction fittings are presented in *Supporting* Figures S5c-f. Such systems might be potentially ambiguous since the interfacial DMI was calculated by first-principle to be quite large (~0.84 mJ/m$^2$)[42], yet a consensus about the magnetic texture identity corresponding to the Hall-humps in different $SrRuO_3$ heterostructures has not been reached via imaging[22a]. Here, the observed γ~0.6 may be inspiring to understand that the $SrRuO_3/BaTiO_3$ heterostructures are in the dipolar-dominant bubbles regime, since its tetragonal crystal structure contributes a large $K_U$. Without the in-plane field schemes (highlighted in our work) to scrutinize the textures' movement, expansion or annihilation, a general magnetic imaging technique would be difficult to reach a complete comprehension on a particular system, especially when texture size is below the imaging resolution limit, or in-plane magnetic field components are inaccessible.

In conclusions, the gradual crossover of Hall-humps with origin from KL-AHE to GHE has been elucidated by observing the response of Hall-humps to in-plane magnetic fields to distinguish the behaviours of collinear domains and SkL ground state. Although the phase boundary between metastable Skyrmions and large collinear domains cannot be identified by our angle-rotation Hall measurement scheme in the range of $0 \ll \gamma < 1$, the presence of small-field SkL can be confirmed by $\gamma \leq 0$ without magnetic imaging. Our work holds promise as an indispensable protocol in complementing magnetic imaging techniques for future development of skyrmion-based topological spintronics and stochastic computing.

**Figures:**

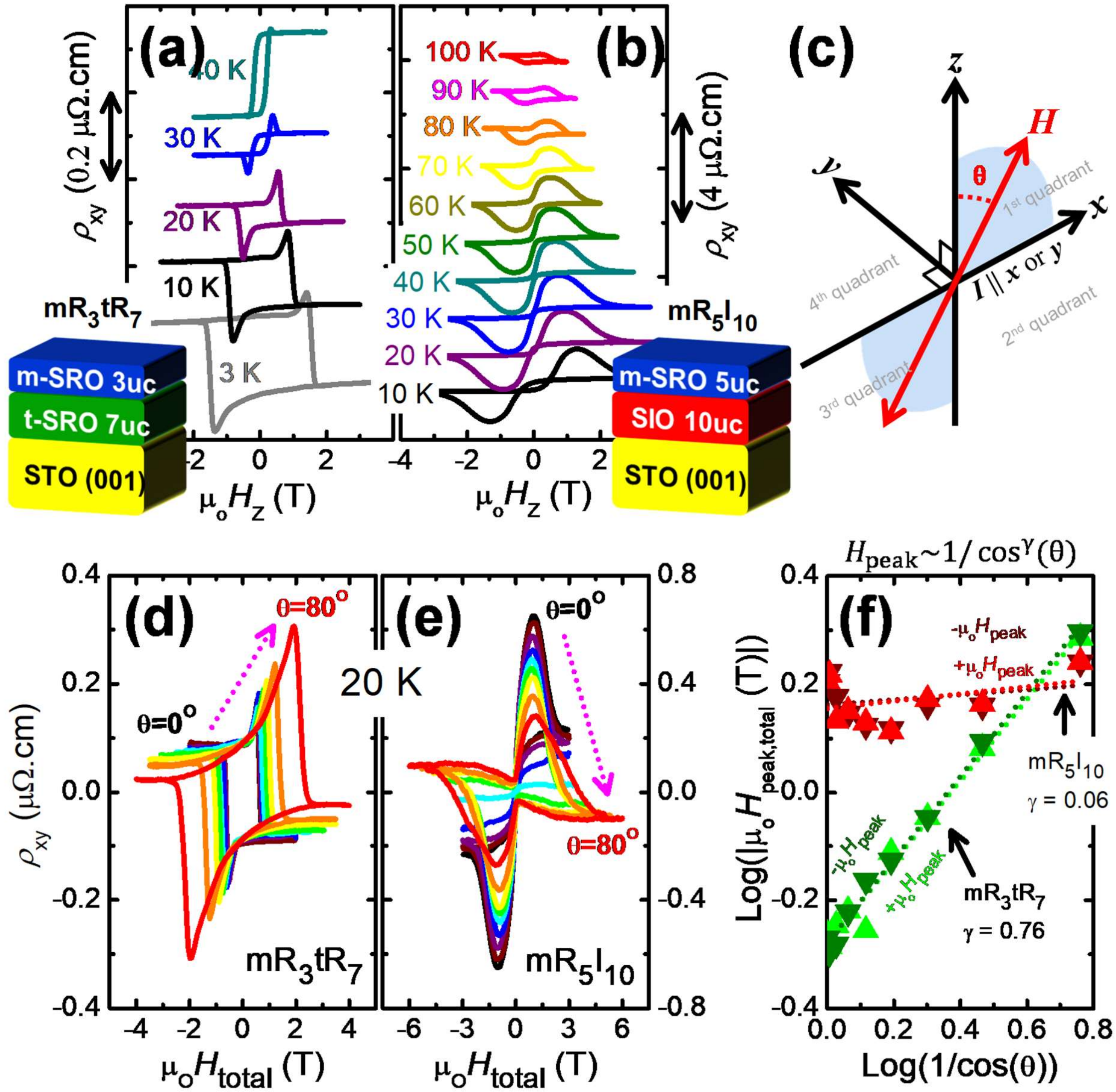

**Figure 1:** Temperature-dependent Hall loops of **(a)** $mR_3tR_7$ and **(b)** $mR_5I_{10}$ on STO(001) substrates with obvious Hall-humps. All OHE components were removed by linear background substraction and loops are shifted vertically. **(c)** Schematic of the total field rotation $\rho_{xy}(H_{\text{total}}, \theta)$ measurements, yielding data for **(d)** $mR_3tR_7$ and **(e)** $mR_5I_{10}$ in the xz-plane at 20 K. **(f)** Log-log plots of $H_{\text{peak}}(\theta)$ originated from (d) and (e), where dotted lines are best-fits for γ extraction.

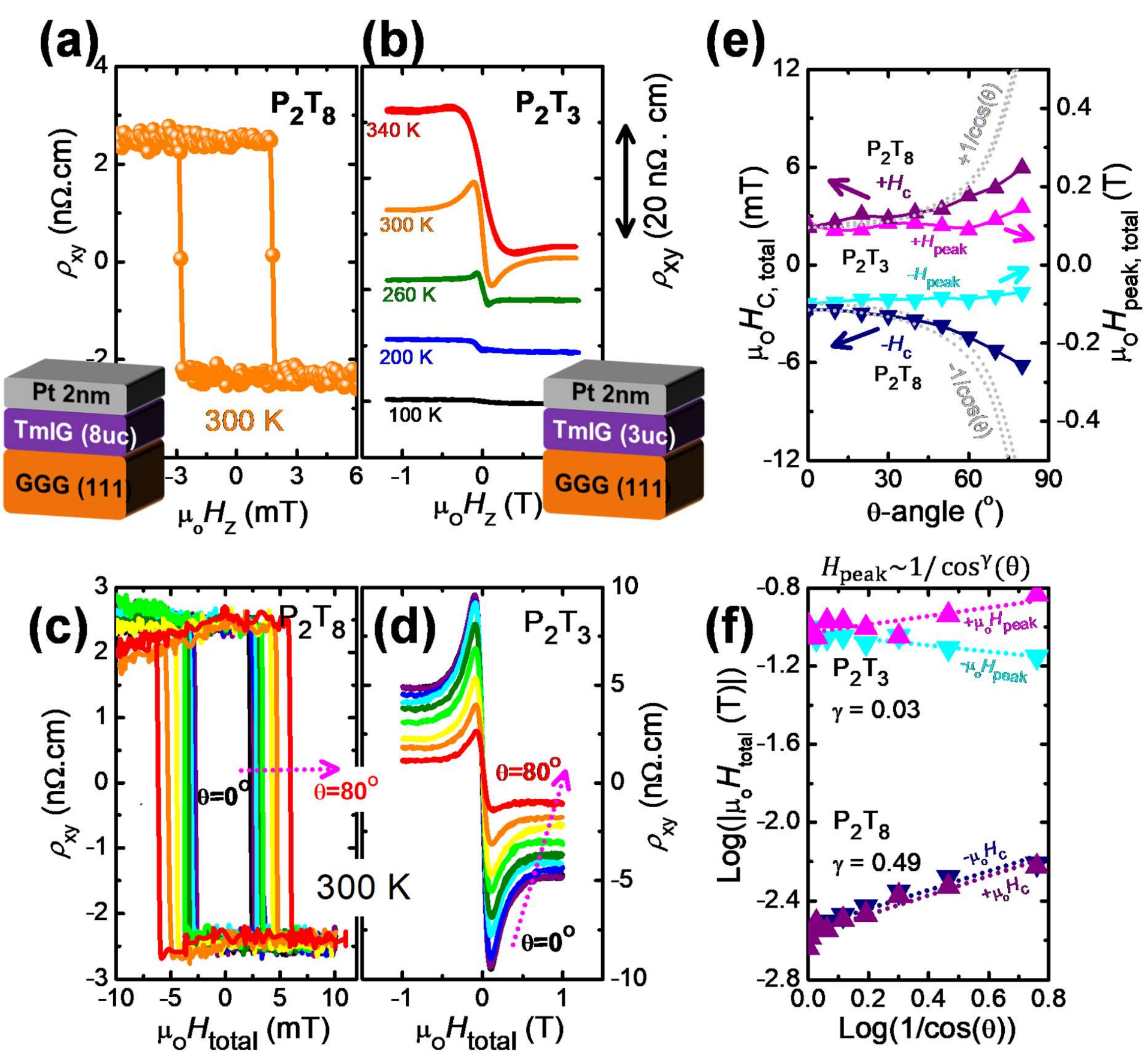

**Figure 2:** Pt-capped TmIG films on GGG(111) substrates with **(a)** 8uc (9.6nm) and **(b)** 3uc (3.6nm) TmIG, labelled as $P_2T_8$ and $P_2T_3$ respectively, for a consistent comparison with the SRO-based heterostructures (main text). $\rho_{xy}(H_{\text{total}}, \theta)$ data for **(c)** $P_2T_8$ and **(d)** $P_2T_3$ under the "total field rotation scheme". (**e**) Field-angle divergence plots and (**f**) Log-log plots for coercive fields of $P_2T_8$ and hump peak fields of $P_2T_3$. Dotted lines in (**e**) are the $1/\cos^1(\theta)$ trends as guide-for-the-eyes to illustrate the data curves have $\gamma<1$, while those in (**f**) are best fits for $\gamma$ extraction.

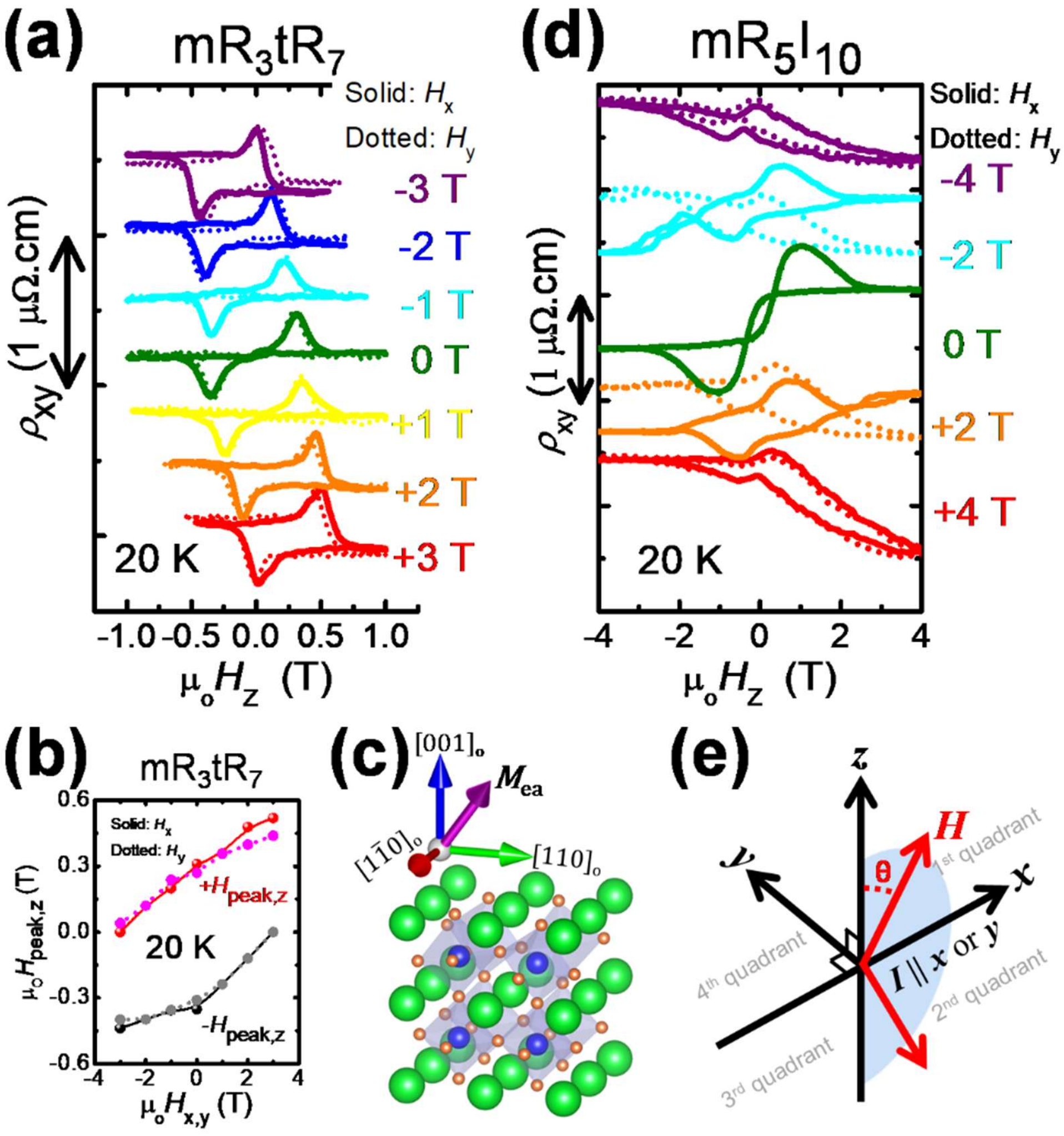

**Figure 3:** $\rho_{xy}(H_\mathrm{z}, H_\mathrm{x,y})$ data for **(a)** $mR_3tR_7$ and **(d)** $mR_5I_{10}$, with comparison between $H_\mathrm{x}$ (solid) and $H_\mathrm{y}$ (dotted) in-plane fields. Loops are vertically shifted for clarity. **(b)** Summary of z-component $H_\mathrm{peak}$ variations with $H_\mathrm{x,y}$ in $mR_3tR_7$ from (a). A low current density $\boldsymbol{J}_\mathrm{c,x}$ of $+6.24\mathrm{x}10^8$ A/m$^2$ was used for both heterostructures. **(c)** Inclined easy axis magnetization ($\boldsymbol{M}_\mathrm{ea}$) of SRO thin film typically occurs on STO(001) substrate. **(e)** Schematic of the resolved field components $\rho_{xy}(H_\mathrm{z}, H_\mathrm{x,y})$ measurements by simultaneous variations of $H_\mathrm{total}$ and θ to maintain a particular $H_\mathrm{x,y}$.

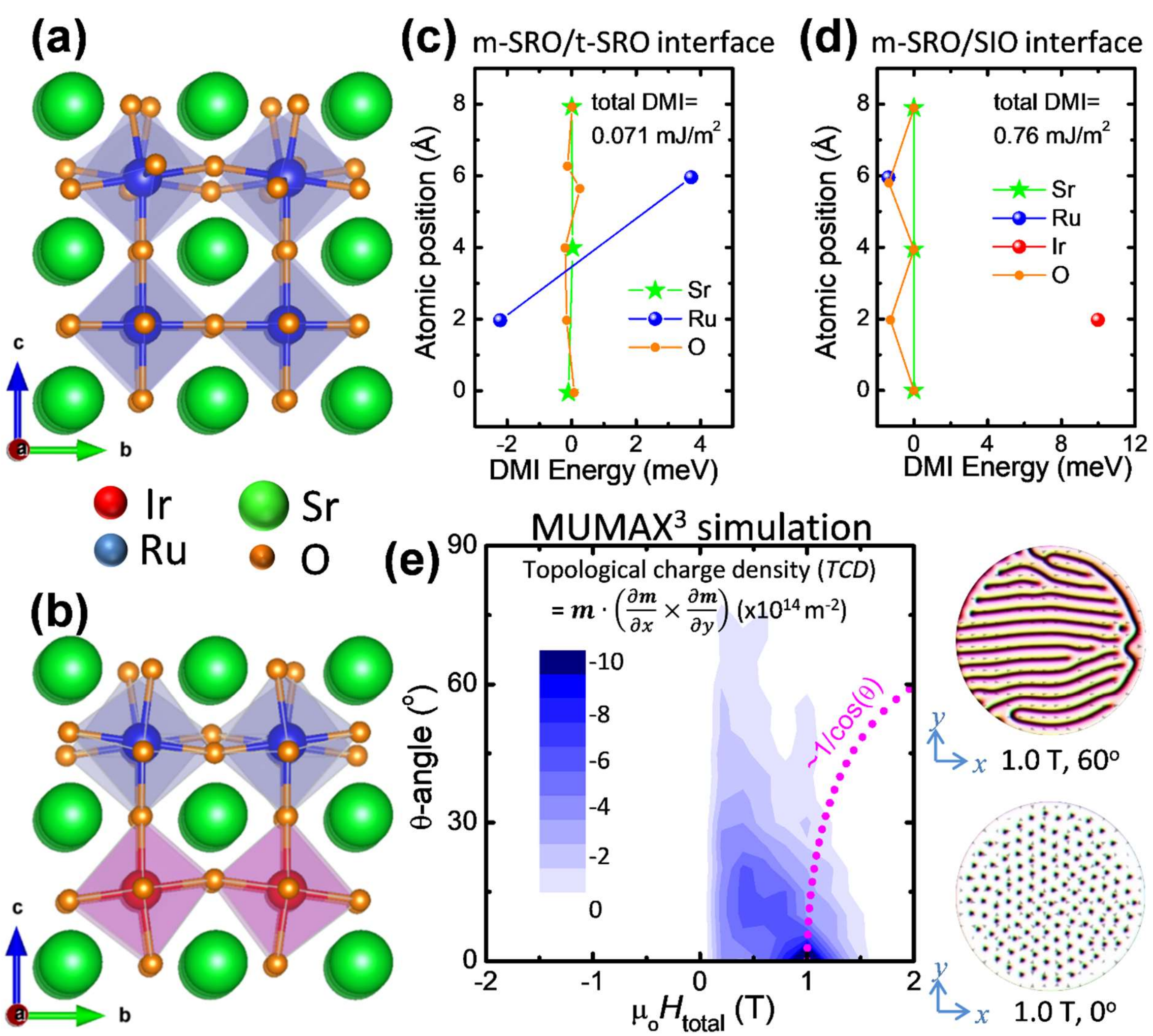

**Figure 4:** DFT structural models for **(a)** $mR_3tR_7$ and **(b)** $mR_5I_{10}$ respectively. Layer and atomic species resolved DMI energies from DFT calculations for **(c)** $mR_3tR_7$ and **(d)** $mR_5I_{10}$ respectively. **(e)** $TCD(H_{total},\theta)$ mapping obtained by MUMAX$^3$ using realistic parameters of $mR_5I_{10}$. Right panels: snapshots of magnetic textures extracted from the $(H_{total},\theta)$ states indicated.

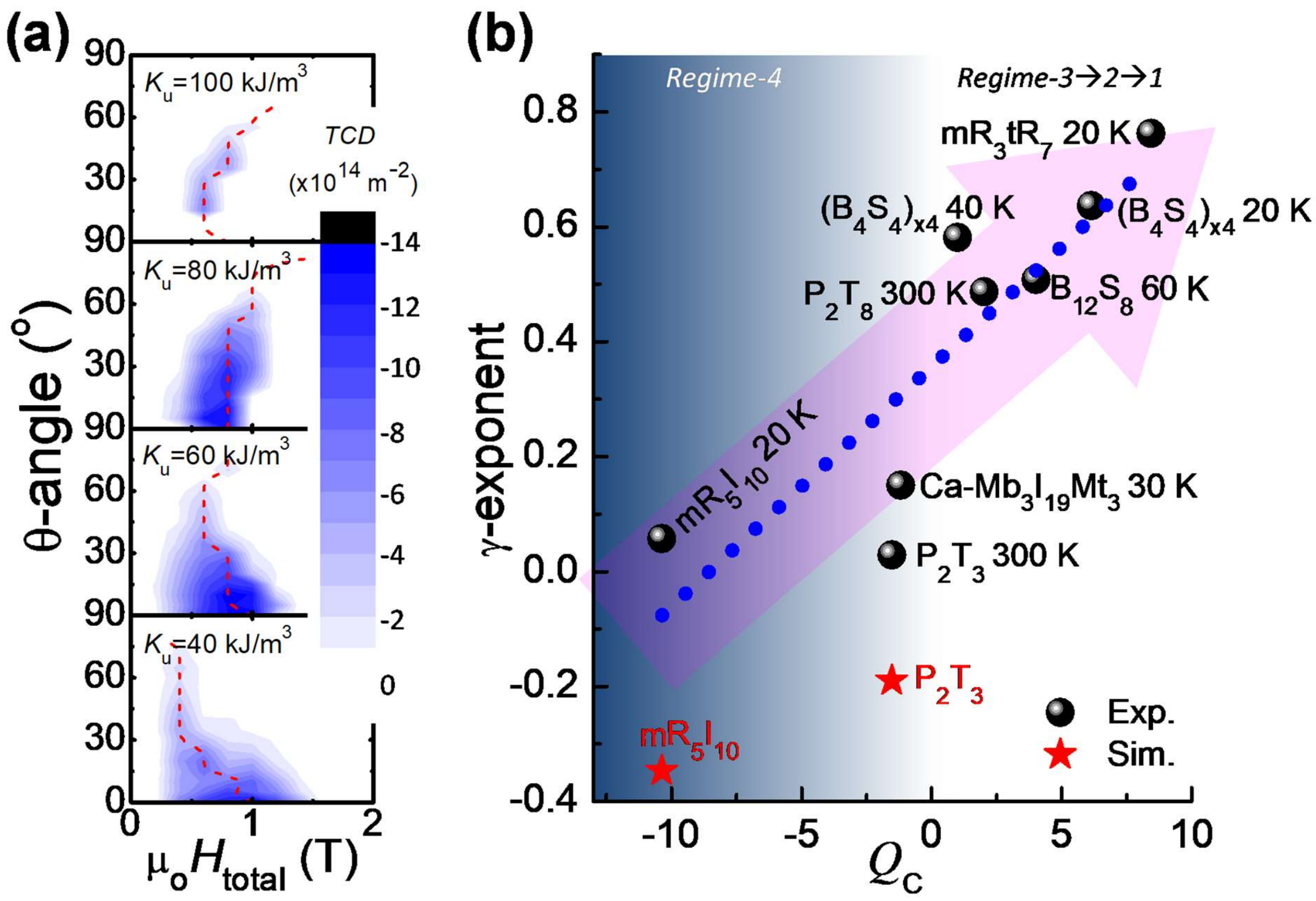

**Figure 5: (a)** *TCD*($H_{\text{total}}$,θ) mappings with varying $K_u$, with the *TCD* peak fields labelled by red dashed curves. **(b)** Trend of γ versus $Q_C$, with parameters extracted from experiments (this work) and various references[42-43]. The final expression of $Q_c = \frac{2K}{\mu_0 M_{\text{sat}}^2} - \frac{\pi|D|}{2\mu_0 M_{\text{sat}}^2}\sqrt{\frac{K}{A_{\text{ex}}}}$. Regime-1: colllinear domains, *Regime-2*: isolated metastable Skyrmions, *Regime-3*: bubble-lattice, *Regime-4*: Néel-type SkL ground state. Details of data are given in *Supporting* Table S2.

**Experimental Section/Methods:**

For all-oxide heterostructures incluing $mR_3tR_7$, $mR_5I_{10}$, $B1_2S_8$ and $(B_4S_4)_{x4}$-SL, the STO(001) substrates' surface were etched by buffered-HF solutions and annealed at 950 °C for 1.5 hours to obtain the single $TiO_2$-termination. Hall-bars were defined by depositing insulating amorphous AlN film with thickness ~200 nm on the STO(001) substrates, followed by photolithography and lift-off. Such Hall-bars fabrication before crystal film growth has the advantages of avoiding ion milling that might induce unwanted oxygen vacancies, water leaching problem[44], or chemical doping[45] during photolithography. Single-crystalline m-phase SRO and SIO films were grown on the exposed STO surface at 660 °C and 750 °C temperatures respectively and oxygen pressure $PO_2$=100 mTorr in a Pulsed Laser Deposition (PLD) system, hence the m-SRO thick film (34uc) shows negative-sign KL-AHE (Figure S1a). Whereas the t-phase SRO was grown at 660 °C and $PO_2$=15 mTorr, showing positive-sign KL-AHE (Figure S1b). The $B_{12}S_8$ was constructed by 8.5uc of SRO film grown on STO(001) at 660 °C and $PO_2$=100 mTorr, followed by 12uc BTO at $PO_2$=2 mTorr and same temperature. The extra 0.5uc of SRO is to account for termination change from $TiO_2$ to SrO at the first uc. Whereas the $(B_4S_4)_{x4}$-SL is actually a superlattice of $[SRO(4uc)$ on $BTO(4uc)]_{x4}$ capped with a final 4uc BTO, constructed by alternating BTO at $PO_2$=10 mTorr and SRO at $PO_2$=50 mTorr at a common temperature 700 °C. The materials deposited on AlN are amorphous and insulating. For structures $P_2T_8$ and $P_2T_3$, the GGG(111) substrates which were annealed at 1000 °C and 6 hours. Then, crystalline TmIG films were grown directly on the substrates at 700 °C and $PO_2$=15 mTorr, before *in-situ* transfer to a sputtering chamber for amorphous Pt capping at room temperature at low speed of 1.62 nm/min to ensure high quality interface. Finally, Hall-bars were defined by photolithography and ion beam milling, since the AlN pre-deposited Hall-bars do not work in cases with Pt capping which is still conducting on amorphous AlN. All Hall-bars are 200-μm long and 20-μm wide. All films have minimal surface roughness of ~rms. 0.2 nm without island growth (step-flow and layer-by-layer for SRO and SIO respectively), so that high quality interfaces are ensured. The final topography of all samples were checked by ParkSystems model NX10 Atomic Force Microscope (AFM). Electrical measurements were done in a Quantum Design (QD) Physical Properties Measurement System (PPMS-6000) equipped with a rotator attachment, while and magnetometry was measured in a QD Superconducting Quantum Interference Device (SQUID) Vibrating Sample Magnetometery (VSM). The PPMS and SQUID-VSM can cool samples down to 2 K and apply magnetic field up to +/-9 T and +/-7 T respectively. In PPMS, before measurements on $P_2T_8$ with low $\mu_o H_C$ of only ~2.2 mT, the magnetic field was first set to 1 T before ramping down to zero in the oscillating mode to remove any residual field due to vortices in the superconducting coil of PPMS. The "total field rotation measurements" for $\rho_{xy}(H_{\text{total}}, \theta)$ were done with pseudoAC currents to eliminate thermal noise, while the "resolved field components measurements" for $\rho_{xy}(H_{\text{z}}, H_{\text{x}})$ were done with DC currents supplied by a Keithley 2400 sourcemeter. X-ray diffraction was measured at the XDD beamline of Singapore Synchrotron Light Source (SSLS). Magnetic imaging by change in frequency ($\Delta f$) was performed by the low-temperature MFM mode of AttoAFM1, which is capable of cooling down to 1.5 K and applying magnetic field up to +/-8 T.

**Supporting Information:**

Supporting Information is available from the Wiley Online Library or from the author.

**Acknowledgements:**

*Authors' contributions:*

A. A. provided all necessary laboratory facilities. Z. S. L. and A.A. constructed the idea and experiment plans. Z. S. L., L. E. C. and G. J. O. grew the oxide films in a PLD system. A. K. H. K. and R. L. performed the DFT calculations. Z. S. L., L. E. C., G. J. O., Z. L, Z. Z. and H. Y. participated in Hall measurements, data analyses and discussions. P. Y. provided valuable experience and guidance at the SSLS-XDD beamline.

*Funding source:*

1. Agency for Science, Technology and Research (A*STAR) under its Advanced Manufacturing and Engineering (AME) Individual Research Grant (IRG) (A1983c0034)
2. National University of Singapore (NUS) Academic Research Fund (A-0004196-00-00)
3. The Singapore National Research Foundation (NRF) under the Competitive Research Programs (CRP Award No. NRF-CRP15-2015-01)
4. NUS-SSLS Core Support (C-380-003-003-001)

*Special Thanks:*

The authors appreciate Max Hirschberger for an insightful short discussion.

**References:**

[1] R. Karplus, J. M. Luttinger, *Physical Review* **1954**, 95, 1154.
[2] a) T. Jungwirth, Q. Niu, A. H. MacDonald, *Phys. Rev. Lett.* **2002**, 88, 207208; b) M. Onoda, N. Nagaosa, *J. Phys. Soc. Jpn.* **2002**, 71, 19.
[3] J. Smit, *Phy* **1958**, 24, 39.
[4] a) L. Berger, *Phys. Rev. B* **1970**, 2, 4559; b) P. Nozières, C. Lewiner, *J. Phys. (Paris)* **1973**, 34, 901.
[5] N. Nagaosa, J. Sinova, S. Onoda, A. H. MacDonald, N. P. Ong, *Reviews of Modern Physics* **2010**, 82, 1539.
[6] a) S. H. Chun, M. B. Salamon, Y. Lyanda-Geller, P. M. Goldbart, P. D. Han, *Phys. Rev. Lett.* **2000**, 84, 757; b) J. Ye, Y. B. Kim, A. J. Millis, B. I. Shraiman, P. Majumdar, Z. Tešanović, *Phys. Rev. Lett.* **1999**, 83, 3737.
[7] a) K. Everschor-Sitte, M. Sitte, *Journal of Applied Physics* **2014**, 115, 172602; b) P. Bruno, V. K. Dugaev, M. Taillefumier, *Phys. Rev. Lett.* **2004**, 93, 096806.
[8] a) Y. Zhang, Y. Sun, H. Yang, J. Železný, S. P. P. Parkin, C. Felser, B. Yan, *Phys. Rev. B* **2017**, 95, 075128; b) H. Ishizuka, N. Nagaosa, *Sci. Adv.* **2018**, 4, eaap9962.
[9] a) I. Dzyaloshinsky, *Journal of Physics and Chemistry of Solids* **1958**, 4, 241; b) T. Moriya, *Physical Review* **1960**, 120, 91.
[10] a) A. O. Leonov, M. Mostovoy, *Nat. Commun.* **2015**, 6, 8275; b) T. Kurumaji, T. Nakajima, M. Hirschberger, A. Kikkawa, Y. Yamasaki, H. Sagayama, H. Nakao, Y. Taguchi, T.-h. Arima, Y. Tokura, *Science* **2019**, 365, 914.
[11] N. Nagaosa, Y. Tokura, *Nat. Nanotechnol.* **2013**, 8, 899.
[12] S. Heinze, K. von Bergmann, M. Menzel, J. Brede, A. Kubetzka, R. Wiesendanger, G. Bihlmayer, S. Blügel, *Nat. Phys.* **2011**, 7, 713.
[13] a) K. Hamamoto, M. Ezawa, N. Nagaosa, *Phys. Rev. B* **2015**, 92, 115417; b) F. R. Lux, F. Freimuth, S. Blügel, Y. Mokrousov, *Communications Physics* **2018**, 1, 60.
[14] F. D. M. Haldane, *Physical Review Letters* **1988**, 61, 2015.
[15] S. Li, Q. Li, J. Liu, N. Ran, P. Lai, L. Shen, J. Xia, L. Xie, Y. Zhou, G. Zhao, *Phys. Rev. B* **2023**, 107, 014414.
[16] K. S. Denisov, I. V. Rozhansky, N. S. Averkiev, E. Lähderanta, *Physical Review B* **2018**, 98, 195439.
[17] a) J. Matsuno, N. Ogawa, K. Yasuda, F. Kagawa, W. Koshibae, N. Nagaosa, Y. Tokura, M. Kawasaki, *Sci. Adv.* **2016**, 2, e1600304; b) D. J. Groenendijk, C. Autieri, T. C. van Thiel, W. Brzezicki, J. R. Hortensius, D. Afanasiev, N. Gauquelin, P. Barone, K. H. W. van den Bos, S. van Aert, J. Verbeeck, A. Filippetti, S. Picozzi, M. Cuoco, A. D. Caviglia, *Phys. Rev. Res.* **2020**, 2, 023404; c) W. Wang, M. W. Daniels, Z. Liao, Y. Zhao, J. Wang, G. Koster, G. Rijnders, C.-Z. Chang, D. Xiao, W. Wu, *Nat. Mater.* **2019**, 18, 1054.
[18] a) K. Yasuda, R. Wakatsuki, T. Morimoto, R. Yoshimi, A. Tsukazaki, K. S. Takahashi, M. Ezawa, M. Kawasaki, N. Nagaosa, Y. Tokura, *Nat. Phys.* **2016**, 12, 555; b) C. Liu, Y. Zang, W. Ruan, Y. Gong, K. He, X. Ma, Q.-K. Xue, Y. Wang, *Phys. Rev. Lett.* **2017**, 119, 176809; c) K. M. Fijalkowski, M. Hartl, M. Winnerlein, P. Mandal, S. Schreyeck, K. Brunner, C. Gould, L. W. Molenkamp, *Phys. Rev. X* **2020**, 10, 011012; d) J. Jiang, D. Xiao, F. Wang, J.-H. Shin, D. Andreoli, J. Zhang, R. Xiao, Y.-F. Zhao, M. Kayyalha, L. Zhang, K. Wang, J. Zang, C. Liu, N. Samarth, M. H. W. Chan, C.-Z. Chang, *Nat. Mater.* **2020**, DOI: 10.1038/s41563-020-0605-z.
[19] a) L. Vistoli, W. Wang, A. Sander, Q. Zhu, B. Casals, R. Cichelero, A. Barthélémy, S. Fusil, G. Herranz, S. Valencia, R. Abrudan, E. Weschke, K. Nakazawa, H. Kohno, J. Santamaria, W. Wu, V. Garcia, M. Bibes, *Nat. Phys.* **2019**, 15, 67; b) K.-Y. Meng, A.

S. Ahmed, M. Baćani, A.-O. Mandru, X. Zhao, N. Bagués, B. D. Esser, J. Flores, D. W. McComb, H. J. Hug, F. Yang, *Nano Lett.* **2019**, 19, 3169.

[20] a) E. Skoropata, J. Nichols, J. M. Ok, R. V. Chopdekar, E. S. Choi, A. Rastogi, C. Sohn, X. Gao, S. Yoon, T. Farmer, R. D. Desautels, Y. Choi, D. Haskel, J. W. Freeland, S. Okamoto, M. Brahlek, H. N. Lee, *Sci. Adv.* **2020**, 6, eaaz3902; b) S. Zhang, F. Kronast, G. van der Laan, T. Hesjedal, *Nano Lett.* **2018**, 18, 1057.

[21] L. Wang, Q. Feng, H. G. Lee, E. K. Ko, Q. Lu, T. W. Noh, *Nano Lett.* **2020**, 20, 2468.

[22] a) S. D. Seddon, D. E. Dogaru, S. J. R. Holt, D. Rusu, J. J. P. Peters, A. M. Sanchez, M. Alexe, *Nat. Commun.* **2021**, 12, 2007; b) Y. Tokura, N. Kanazawa, *Chemical Reviews* **2021**, 121, 2857.

[23] Q. L. He, G. Yin, A. J. Grutter, L. Pan, X. Che, G. Yu, D. A. Gilbert, S. M. Disseler, Y. Liu, P. Shafer, B. Zhang, Y. Wu, B. J. Kirby, E. Arenholz, R. K. Lake, X. Han, K. L. Wang, *Nat. Commun.* **2018**, 9, 2767.

[24] a) A. M. Glazer, *AcCrA* **1975**, 31, 756; b) A. M. Glazer, *AcCrB* **1972**, 28, 3384.

[25] W. Lu, P. Yang, W. D. Song, G. M. Chow, J. S. Chen, *Phys. Rev. B* **2013**, 88, 214115.

[26] a) Q. Qin, L. Liu, W. Lin, X. Shu, Q. Xie, Z. Lim, C. Li, S. He, G. M. Chow, J. Chen, *Adv. Mater.* **2019**, 31, 1807008; b) D. Kan, T. Moriyama, K. Kobayashi, Y. Shimakawa, *Phys. Rev. B* **2018**, 98, 180408.

[27] Z. Liao, M. Huijben, Z. Zhong, N. Gauquelin, S. Macke, R. J. Green, S. Van Aert, J. Verbeeck, G. Van Tendeloo, K. Held, G. A. Sawatzky, G. Koster, G. Rijnders, *Nat. Mater.* **2016**, 15, 425.

[28] a) C. O. Avci, A. Quindeau, C.-F. Pai, M. Mann, L. Caretta, A. S. Tang, M. C. Onbasli, C. A. Ross, G. S. D. Beach, *Nat. Mater.* **2017**, 16, 309; b) C. O. Avci, E. Rosenberg, L. Caretta, F. Büttner, M. Mann, C. Marcus, D. Bono, C. A. Ross, G. S. D. Beach, *Nature Nanotechnology* **2019**, 14, 561; c) S. Vélez, J. Schaab, M. S. Wörnle, M. Müller, E. Gradauskaite, P. Welter, C. Gutgsell, C. Nistor, C. L. Degen, M. Trassin, M. Fiebig, P. Gambardella, *Nat. Commun.* **2019**, 10, 4750.

[29] a) A. S. Ahmed, A. J. Lee, N. Bagués, B. A. McCullian, A. M. A. Thabt, A. Perrine, P.-K. Wu, J. R. Rowland, M. Randeria, P. C. Hammel, D. W. McComb, F. Yang, *Nano Lett.* **2019**, 19, 5683; b) Q. Shao, Y. Liu, G. Yu, S. K. Kim, X. Che, C. Tang, Q. L. He, Y. Tserkovnyak, J. Shi, K. L. Wang, *Nat. Electron.* **2019**, 2, 182; c) A. J. Lee, A. S. Ahmed, J. Flores, S. Guo, B. Wang, N. Bagués, D. W. McComb, F. Yang, *Phys. Rev. Lett.* **2020**, 124, 107201.

[30] M. Raju, A. P. Petrović, A. Yagil, K. S. Denisov, N. K. Duong, B. Göbel, E. Şaşıoğlu, O. M. Auslaender, I. Mertig, I. V. Rozhansky, C. Panagopoulos, *Nat. Commun.* **2021**, 12, 2758.

[31] L. Liu, Q. Qin, W. Lin, C. Li, Q. Xie, S. He, X. Shu, C. Zhou, Z. Lim, J. Yu, W. Lu, M. Li, X. Yan, S. J. Pennycook, J. Chen, *Nat. Nanotechnol.* **2019**, 14, 939.

[32] T. C. van Thiel, J. Fowlie, C. Autieri, N. Manca, M. Šiškins, D. Afanasiev, S. Gariglio, A. D. Caviglia, *ACS Materials Letters* **2020**, 2, 389.

[33] A. Vansteenkiste, J. Leliaert, M. Dvornik, M. Helsen, F. Garcia-Sanchez, B. V. Waeyenberge, *AIP Advances* **2014**, 4, 107133.

[34] a) A. O. Leonov, I. Kézsmárki, *Phys. Rev. B* **2017**, 96, 214413; b) X. Wan, Y. Hu, B. Wang, *Phys. Rev. B* **2019**, 99, 180406; c) S. Zhang, J. Zhang, Y. Wen, E. M. Chudnovsky, X. Zhang, *Communications Physics* **2018**, 1, 36.

[35] A. Bogdanov, A. Hubert, *Journal of Magnetism and Magnetic Materials* **1994**, 138, 255.

[36] I. Kézsmárki, S. Bordács, P. Milde, E. Neuber, L. M. Eng, J. S. White, H. M. Rønnow, C. D. Dewhurst, M. Mochizuki, K. Yanai, H. Nakamura, D. Ehlers, V. Tsurkan, A. Loidl, *Nat. Mater.* **2015**, 14, 1116.

[37] T. Kurumaji, T. Nakajima, V. Ukleev, A. Feoktystov, T.-h. Arima, K. Kakurai, Y. Tokura, *Phys. Rev. Lett.* **2017**, 119, 237201.

[38] a) J. A. Cape, G. W. Lehman, **1971**, 42, 5732; b) N. S. Kiselev, A. N. Bogdanov, R. Schäfer, U. K. Rößler, *J. Phys. D: Appl. Phys.* **2011**, 44, 392001; c) I. Lemesh, F. Büttner, G. S. D. Beach, *Phys. Rev. B* **2017**, 95, 174423.
[39] H. Zhang, D. Raftrey, Y.-T. Chan, Y.-T. Shao, R. Chen, X. Chen, X. Huang, J. T. Reichanadter, K. Dong, S. Susarla, L. Caretta, Z. Chen, J. Yao, P. Fischer, J. B. Neaton, W. Wu, D. A. Muller, R. J. Birgeneau, R. Ramesh, *Sci. Adv.* **2022**, 8, eabm7103.
[40] S. Parkin, S.-H. Yang, *Nat. Nanotechnol.* **2015**, 10, 195.
[41] J. Iwasaki, M. Mochizuki, N. Nagaosa, *Nat. Commun.* **2013**, 4, 1463.
[42] L. Wang, Q. Feng, Y. Kim, R. Kim, K. H. Lee, S. D. Pollard, Y. J. Shin, H. Zhou, W. Peng, D. Lee, W. Meng, H. Yang, J. H. Han, M. Kim, Q. Lu, T. W. Noh, *Nat. Mater.* **2018**, 17, 1087.
[43] a) L. Caretta, E. Rosenberg, F. Büttner, T. Fakhrul, P. Gargiani, M. Valvidares, Z. Chen, P. Reddy, D. A. Muller, C. A. Ross, G. S. D. Beach, *Nat. Commun.* **2020**, 11, 1090; b) C. Tang, P. Sellappan, Y. Liu, Y. Xu, J. E. Garay, J. Shi, *Phys. Rev. B* **2016**, 94, 140403; c) Z. S. Lim, C. Li, Z. Huang, X. Chi, J. Zhou, S. Zeng, G. J. Omar, Y. P. Feng, A. Rusydi, S. J. Pennycook, T. Venkatesan, A. Ariando, *Small* **2020**, 16, 2004683.
[44] G. Kimbell, C. Kim, W. Wu, M. Cuoco, J. W. A. Robinson, *Communications Materials* **2022**, 3, 19.
[45] Z. Li, S. Shen, Z. Tian, K. Hwangbo, M. Wang, Y. Wang, F. M. Bartram, L. He, Y. Lyu, Y. Dong, G. Wan, H. Li, N. Lu, J. Zang, H. Zhou, E. Arenholz, Q. He, L. Yang, W. Luo, P. Yu, *Nat. Commun.* **2020**, 11, 184.

# Table of Content

Introductions: there are ambiguities between magnetic textures with step-function-like profile and continuous sinusoidal profile from the textures' centre to the surroundings.

Results: We compare the angular dependence behaviour between two distinct groups of samples showing hump-shape Hall Effect. We can distinguish between the Karplus-Luttinger type corresponding to collinear domains and the Geometrical type corresponding to magnetic Skyrmion lattice.

Discussions: We constructed a scaling analysis of Hall-hump peak field divergence exponent versus domain wall energies among various samples. It is a powerful tool to clarify the evolution of the mentioned profile and complement the shortcomings of general imaging results.

**Angular dependence of hump-shape Hall Effects for distinguishing between Karplus-Luttinger and Geometrical Origins**

*Zhi Shiuh Lim*[1,#], *Lin Er Chow*[1], *Amy Khoong Hong Khoo*[2], *Ganesh Ji Omar*[1], *Zhaoyang Luo*[1], *Zhaoting Zhang*[1], *Hong Yan*[1], *Ping Yang*[3], *Robert Laskowski*[2], *A. Ariando*[1,*]

ToC figure: ((Please choose one size: 55 mm broad × 50 mm high **or** 110 mm broad × 20 mm high. Please do not use any other dimensions))

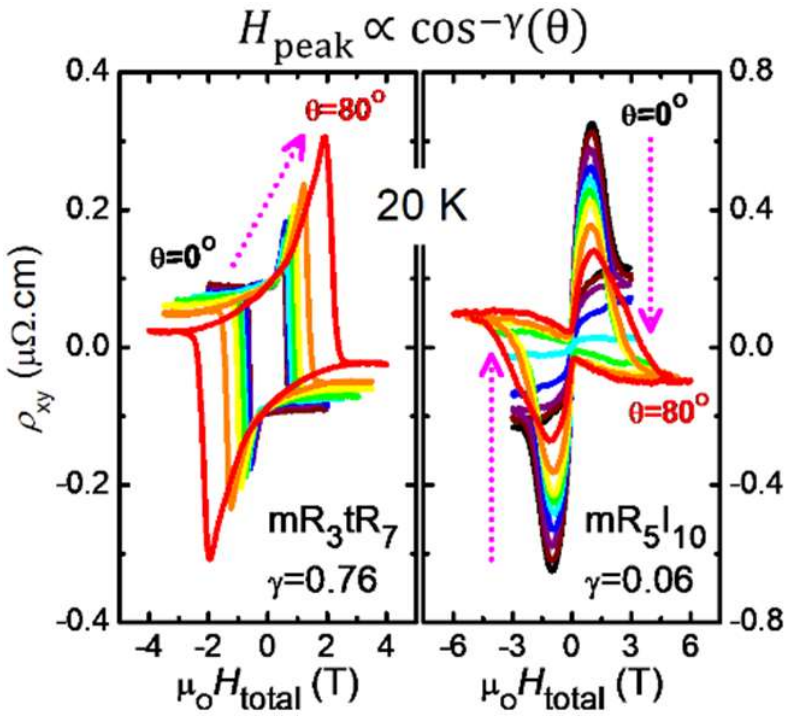

# Supporting Information

**Angular dependence of hump-shape Hall Effects for distinguishing between Karplus-Luttinger and Geometrical Origins**

*Zhi Shiuh Lim[1,#], Lin Er Chow[1], Amy Khoong Hong Khoo[2], Ganesh Ji Omar[1], Zhaoyang Luo[1], Zhaoting Zhang[1], Hong Yan[1], Ping Yang[3], Robert Laskowski[2], A. Ariando[1,*]*

## Additional Analyses on X-ray Diffraction

In Figure S1c, the half-integer XRD peaks of m-SRO and t-SRO thick films (34uc) are shown, with notations and HKL-indices following the Glazer's rules[1]. We believe that the broad humps around (0.5,1.5,1) are measurement artifacts. Considering that bulk m-SRO material is known to belong to the *pbnm* space group with $a^-a^-c^+$ tilt pattern, its in-phase (+) oxygen octahedral rotation axis should have shorter B-O-B bond length. Since the STO(001) substrate imposes an in-plane compressive strain to the film, it is energetically favourable for the in-phase rotation axis of m-SRO to align with the axes "*a*" or "*b*" under compression[2]. Hence, $a^-a^-c^+$ is unlikely to occur in thin film cases before strain relaxation, while $a^-b^+c^-$ and $a^+b^-c^-$ may occur with equal probability[2-3]. In our case, strong peaks HKL=(0.5,0.5,1.5) and (1.5,0.5,1.5), supports the existence of $a^-$. Then, a peak occuring at (0.5,0,1.5) supports $b^+$ and rules out $b^-$, hence $c^-$ can be deduced from (0.5, 1.5,1.5), forming $a^-b^+c^-$. In t-SRO(34uc), most peaks are suppressed, consistent to tetragonal $a^0a^0c^0$ whch is also partially stabilized by oxygen vacancies[4]. Similarly, almost all Half-integer XRD peaks are suppressed in $mR_3tR_7$ due to the dominant t-SRO. Hence, the interface is assigned as SRO($a^-b^+c^0$)/SRO($a^0b^0c^0$) as a reasonable approximation for subsequent DFT calculations.

For $mR_5I_{10}$, the SIO film was deliberately designed to be thick enough for relaxation from strain and tilt-suppression by the cubic STO(001) substrates, hence the $a^+$ and $b^+$ tilts are likely to form in SIO at regions away from the substrate. The peak at (0,0.5,1.5) thus belongs to the $a^+$ of the SIO film (red curve of Figure S1c) at region near the SRO/SIO interface, and can be assigned as $a^+b^0c^0$. Subsequently, the top SRO can be expected to have weak $a^-b^+c^0$, bearing much resemblance to the thick m-SRO (34uc) with $a^-b^+c^-$. Previously in reference[5], we have discussed that the thick enough bottom SIO with relaxation from $a^0b^0c^0$ to $a^+b^0c^0$ is crucial to achieve the Hall-humps. Replacing the bottom SIO with a low pressure grown one with the same thickness will again result in total suppression in almost all half-interger XRD peaks, since oxygen vacancies in SIO also tend to stabilize tetragonal crystal structure, leading to suppression of Hall-humps. We believe the $c^+$ peaks at (0.5,1.5,1.0) visible in m-SRO(34uc), t-SRO(34uc) and $mR_3tR_7$ are resulted from further strain relaxation near the sample/air surface recovering its bulk-like *Pbnm* state of $a^-a^-c^+$, but are far away from and not relevant to the crucial interfaces responsible for DMI highlighted in main text Figure 3a,b.

**Additional Data for Damping-like SOT of $P_2T_8$**

In Figure S3a,b, $P_2T_8$ also exhibited an additional symmetric loop-narrowing effect ($\pm H_{C,z}$ reduction) at the small regimes of $\mu_o|H_x|<5$ mT. This effect can be understood from purely magnetic field torque acting on $\pm M_z$ domains causing easier switching for both polarities and is irrelevant to damping-like SOT on $m_{DW}$. Damping-like SOT becomes more relevant at $\mu_o|H_x|>5$ mT, since larger $H_x$ may enhance the effect of damping-like SOT even at small $J_{c,x}<10^{10}$ A/m$^2$. Such symmetrical $H_z$ loop-narrowing regime at low $\pm H_x$ was also observed but less frequently in different samples of $mR_3tR_7$. Hence, trivial magnetic bubbles can be inferred to be dominant in $mR_3tR_7$ with abundant DWs, while $P_2T_8$ is closer to abrupt single collinear $M_z$ domain switching.

**First-principle Calculations**

In Density Functional Theory (DFT) calculations for Dzyaloshinskii-Moriya Interaction (DMI), we limited the structure in each case to a two-unit cell thin slab for optimal computational cost (main text Figure 3a,b). Before relaxation, the in-plane lattice parameters *a* and *b* are fixed at 3.9053 Å following the substrate, while the out-of-plane lattice parameters *c* = 4.04 Å for t-SRO and 3.94 Å for m-SRO in $mR_3tR_7$. Likewise in $mR_5I_{10}$, the lattice parameters are *a*=*b*=3.9053 Å, while *c* = 3.95 Å for SIO and 3.94 Å for mSRO. Next, the WIEN2k package[6] was employed in three steps [7]. Firstly, structural relaxations were performed on top of the input obtained from half-integer HKL XRD until the forces became smaller than 0.01 eV/Å for determining the most stable interfacial geometries, resulting in small adjustments from the input. Secondly, the Kohn-Sham equations were solved using the generalized Bloch theorem[8] as implemented in non-collinear spin version of WIEN2k (WIENNCM) to generate wavefunctions of spin spiral state. An arbitrary spiral vector q of 0.1x2π/a* was chosen, where a*=2a is the in-plane lattice constant of the supercell needed to account for the octahedral rotation. Finally, a condition for the generalized Bloch theorem to be valid is that spin is isotropic and this precludes the inclusion of spin-orbit interactions. Thus to recover spin-orbit effects, we employ first-order perturbation for the spin-orbit Hamiltonian with wavefunctions obtained for the spin spiral to obtain the spin-orbit energy[9]. This energy is then divided by the system volume and set to Dq to obtain the value of total DMI (main text Figure 3c,d).

Interestingly, the $mR_3tR_7$ sample is found to be unstable in DFT, consistent to experimental observation. When $mR_3tR_7$ was relaxed following the protocol of "forces <0.01 eV/Å" mentioned, the octahedral rotation angle across the interface became vanishingly small. In experiments, we observed a gradual evolution of all $mR_3tR_7$ samples from showing the Hall-humps (main text Figure 1a) into a single positive AHE square loop (hump vanished) within the course of 3-4 days if the sample is stored under ambient condition. This can be understood where the top 3uc m-SRO becomes progressively dominated by the bottom 7uc t-SRO layer at room temperature due to oxygen vacancy diffusion across the interface or atomic position shifting. Hence, all the subsequent magnetic field rotation analyses are measured immediately after fresh sample growth, while the low measurement temperature (20 K) is capable of slowing down the evolution. Hence, in DFT, the DMI calculation of $mR_3tR_7$ was done without structural relaxation, and represents a metastable state. This is in contrast to $mR_5I_{10}$ where the full structural relaxation did not produce significant change to the initial model built from half-integer XRD, while in experiment the Hall-humps of $mR_5I_{10}$ is stable over several years.

**Magnetometry and MUMAX3 Micromagnetic Simulations**

In Fig. S5a-d, the magnetometry data of $mR_3tR_7$, $mR_5I_{10}$, $B1_2S_8$, $(B_4S_4)_{x4}$-SL are shown. The *M-H* loops with $H||$z (out-of-plane) and $H||$x,y (in-plane) at the left panels are useful to calculate magnetic anisotropy and inference of saturation magnetization ($M_{sat}$) at various teemperatures. The Curie temperatures ($T_C$) can be inferred from the moment versus temperature (*M-T*) from the right panels. The field-cooling (FC) *M-T* curves of Figure S4b extending to higher temperature cut-offs (~150 K) than the zero-field-cooled (ZFC) *M-T* curves are likely due to the proximity-induced magnetism and spin-liquid behaviour of SIO film. These information are used as input parameters for MUMAX$^3$ simulation, using the following equations:

$$A_{\mathrm{ex}} = \frac{3k_{\mathrm{B}}T_{\mathrm{c}}}{2j(j+1)a}$$

$$l_{\mathrm{ex}} = \sqrt{\frac{2A_{\mathrm{ex}}}{\mu_{\mathrm{o}}{M_{\mathrm{sat}}}^2}}$$

$$K_{\mathrm{U}} = \int_0^{H_{\mathrm{sat}}} (M_{\boldsymbol{H}||\boldsymbol{c}} - M_{\boldsymbol{H}||\boldsymbol{a}}) \cdot dH + \mu_{\mathrm{o}}{M_{\mathrm{sat}}}^2/2$$

where $A_{ex}$, j, $a$, $l_{ex}$, and $K_U$ are the exchange stiffness, total angular momentum quantum number per formula unit, lattice parameter, exchange length, and magnetocrystalline anisotropy respectively. Since $Ru^{4+}$ ($4d^4$) is lying between the weak SOC (l-s coupling) and strong SOC (j-j coupling) regimes with spin and orbital angular momentum of S=1, L=1, the SRO has j=1 because the two vectors are neither parallel nor antiparallel. While j=5/2 is typically suitable for $Fe^{3+}$ ($3d^5$) with S=5/2 and L=0 in TmIG.

For SRO-based heterostructures, relatively low temperatures were chosen for $K_u$ calculations as indicated in the *M-H* curves to match the fact that the DMI values extracted from DFT are ground state (0 K) values, and we do not know the trend of DMI reduction with thermal fluctuation. This strategy facilitates getting a good trend in main text Figure 4b. Due to presence of $Ir^{4+}$ paramagnetic species in $mR_5I_{10}$, the $M_{sat}$ data obtained from SQUID would be slightly less accurate. Hence, we allowed $M_{sat}$ as the only free variable for manual optimization to match the TCD peak field to the peak of Hall-hump data at $\mu_o H_{peak}$=1.0 T and θ=0, before simulations of other θ-angles. Likewise, due to the difficulty in measuring precise magnetometry data for TmIG thin films grown on highly-paramagnetic GGG(111) substrates, $K_{eff}$ of $P_2T_8$ were extracted from reference[10] showing anisotropy field, $H_K$ = 1460 Oe, and $K_{eff} \approx \mu_o M_{sat} H_K$ =13140 J/m$^3$ while $M_{sat}$ = 100 emu/cm$^3$ = 1x10$^5$ A/m. Whereas $T_C$ of $P_2T_8$ and $P_2T_3$ can be confidently estimated to be 550 K and 300 K respectively, since 550 K is known to be the bulk value of garnet ferrites[11]. Whereas from the Main Text Figure 2b we can see a divergence of Hall-hump (genuine Spin-Hall Geometrical Hall Effect (SH-THE) in $P_2T_3$ with temperature around 300 K, consistent to the power-law scaling of $\rho_{xy}^{T} \propto |T - T_C|^{-\gamma'}$ discussed in reference[12], verifying its lower 2$^{nd}$-order phase transition temperature. A simulation grid cell size of 4 nm < $l_{ex}$ was used for all cases, with 256x256 meshes in the xy-plane but 1-cell in the z-direction thickness. The backward/forward field sweep simulations are always initialized with m=Uniform(0,0,±1) to retain the history of magnetization saturation. Each simulation was terminated after reaching a criterion of max-torque < 5x10$^{-3}$. The simulation parameters for $mR_5I_{10}$ and $P_2T_3$ are shown in the table S1 below, resulting in hysteretic *TCD* in $mR_5I_{10}$ but non-hysteretic TCD in $P_2T_3$.

| | $T_C$ | j | $a$ | $A_{ex}$ | $M_{sat}$ | $K_U$ | $D_{int}$ | $l_{ex}$ | Thickness | $a_G$ |
|---|---|---|---|---|---|---|---|---|---|---|
| Unit | K | | Å | pJ/m | kA/m | kJ/m$^3$ | mJ/m$^2$ | nm | nm | |
| $mR_5I_{10}$ | 125 | 1 | 3.93 | 3.43 | 63 | 24.8 | 0.76 | 36.3 | 1.97 | 0.1 |
| $P_2T_3$ | 300 | 5/2 | 12.3 | 0.58 | 30 | 1 | 0.1 | 47.9 | 3.70 | 0.03 |

**Supporting Table S1:** MUMAX$^3$ Simulation parameters of $mR_5I_{10}$ and $P_2T_3$ for $TCD(H_{total},\theta)$ mappings.

Figure 5b in the main text was generated from the following data of table S2. $K_{eff}$ and $D$ of $P_2T_8$ were extracted from [10] and [13], Ca-$Mb_3I_{19}Mt_3$ from [14], while $K_{eff}$, $A_{ex}$ and $D$ of $B_{12}S_8$ use values of $B_{20}S_5$ from reference [15], since its magnetometry signal is too weak to be measured accurately from our SQUID-VSM setup. The rest of parameter values were experimentally determined in this work.

| **Samples** | $\boldsymbol{A_{ex}}$ (pJ/m) | $\boldsymbol{D}$ (mJ/m$^2$) | $\boldsymbol{M_{sat}}$ (kA/m) | $\boldsymbol{K_{eff}}$ (kJ/m$^3$) | $\boldsymbol{K_U}$ (kJ/m$^3$) | $\boldsymbol{K}$ (mJ/m$^2$) | $\boldsymbol{Q_C}$ | **γ** |
|---|---|---|---|---|---|---|---|---|
| $mR_5I_{10}$ | 3.43 | 0.76 | 63.0 | 22.3 | 24.8 | 22.3 | -10.4 | 0.06 |
| $mR_5I_{10}$ (sim.) | 3.43 | 0.76 | 63.0 | | 24.8 | 22.3 | -10.4 | -0.35 |
| $P_2T_3$ | 0.58 | 0.06 | 30 | 0.43 | 1 | 0.43 | -1.52 | 0.03 |
| $P_2T_3$ (sim.) | 0.58 | 0.06 | 30 | | 1 | 0.43 | -1.52 | -0.19 |
| $P_2T_8$ | 1.08 | 0.0052 | 100 | 13.1 | 19.4 | 13.1 | 2.02 | 0.49 |
| Ca-$Mb_3I_{19}Mt_3$ | 1.23 | 0.42 | 153 | 6.54 | 21.3 | 6.54 | -1.19 | 0.15 |
| $B_{12}S_8$ (60 K) | 3.4 | 0.84 | 61.7 | 146 | 148 | 146 | 4.01 | 0.51 |
| $(B_4S_4)_{x4}$ (20 K) | 3.4 | 0.84 | 102.3 | 200.2 | 206.8 | 200.2 | 6.15 | 0.64 |
| $(B_4S_4)_{x4}$ (40 K) | 3.4 | 0.58 | 89.36 | 70.8 | 75.8 | 70.8 | 1 | 0.58 |
| $mR_3tR_7$ | 3.2 | 0.071 | 135.8 | 107.9 | 119.4 | 107.9 | 8.43 | 0.76 |

**Supporting Table S2:** Data list for the main text Figure 5b's plot.

To clarify:

$$K = \begin{cases} K_{\mathrm{eff}} & , \quad \text{experiments} \\ K_{\mathrm{U}} - \frac{\mu_{\mathrm{o}} M_{\mathrm{sat}}{}^{2}}{2} & , \quad \text{simulations} \end{cases}$$

Subsequently,

$$w = \sqrt{\frac{A_{\mathrm{ex}}}{K}}$$

$$\sigma_{\mathrm{DW}} = 4\sqrt{A_{\mathrm{ex}}K} - \pi|D|$$

$$l_{\mathrm{d}} = \frac{\sigma_{\mathrm{DW}}}{4K_{\mathrm{shape}}} = \frac{2\sqrt{A_{\mathrm{ex}}K} - \pi|D|/2}{\mu_{\mathrm{o}} M_{\mathrm{sat}}^{2}}$$

$$Q_{c} = \frac{l_{\mathrm{d}}}{w} = \frac{2K}{\mu_{\mathrm{o}} M_{\mathrm{sat}}^{2}} - \frac{\pi|D|}{2\mu_{\mathrm{o}} M_{\mathrm{sat}}^{2}}\sqrt{\frac{K}{A_{\mathrm{ex}}}}$$

**Supporting Figures:**

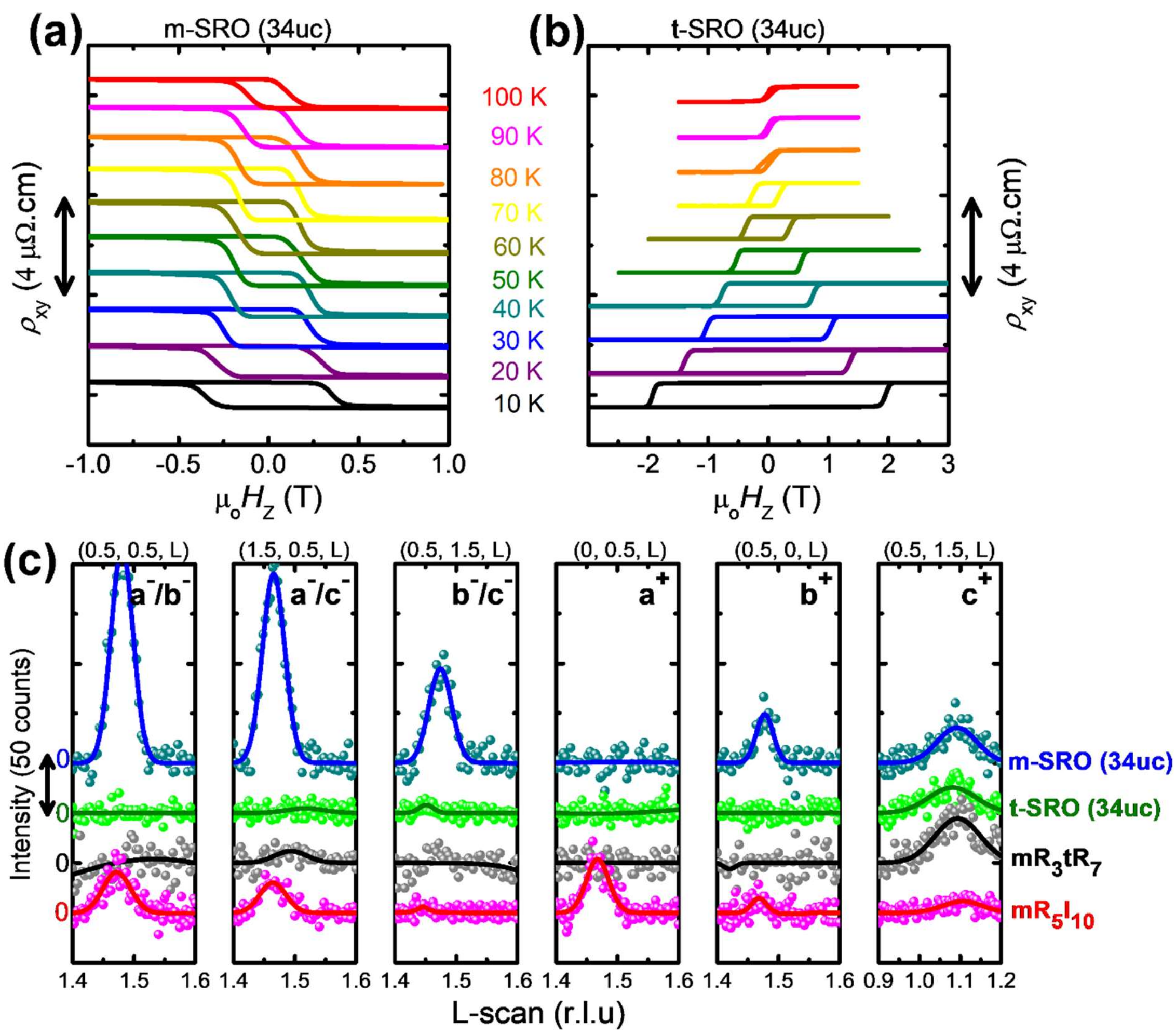

**Supporting Figure S1:** KL-AHE loops of **(a)** m-phase SRO and **(b)** t-phase SRO thick films (34uc each) on STO(001) substrates without Hall-humps. Data of different temperatures are vertically shifted for clarity. **(b)** Half-integer X-ray Bragg diffraction data for thick films of m-SRO and t-SRO, $mR_3tR_7$ and $mR_5I_{10}$.

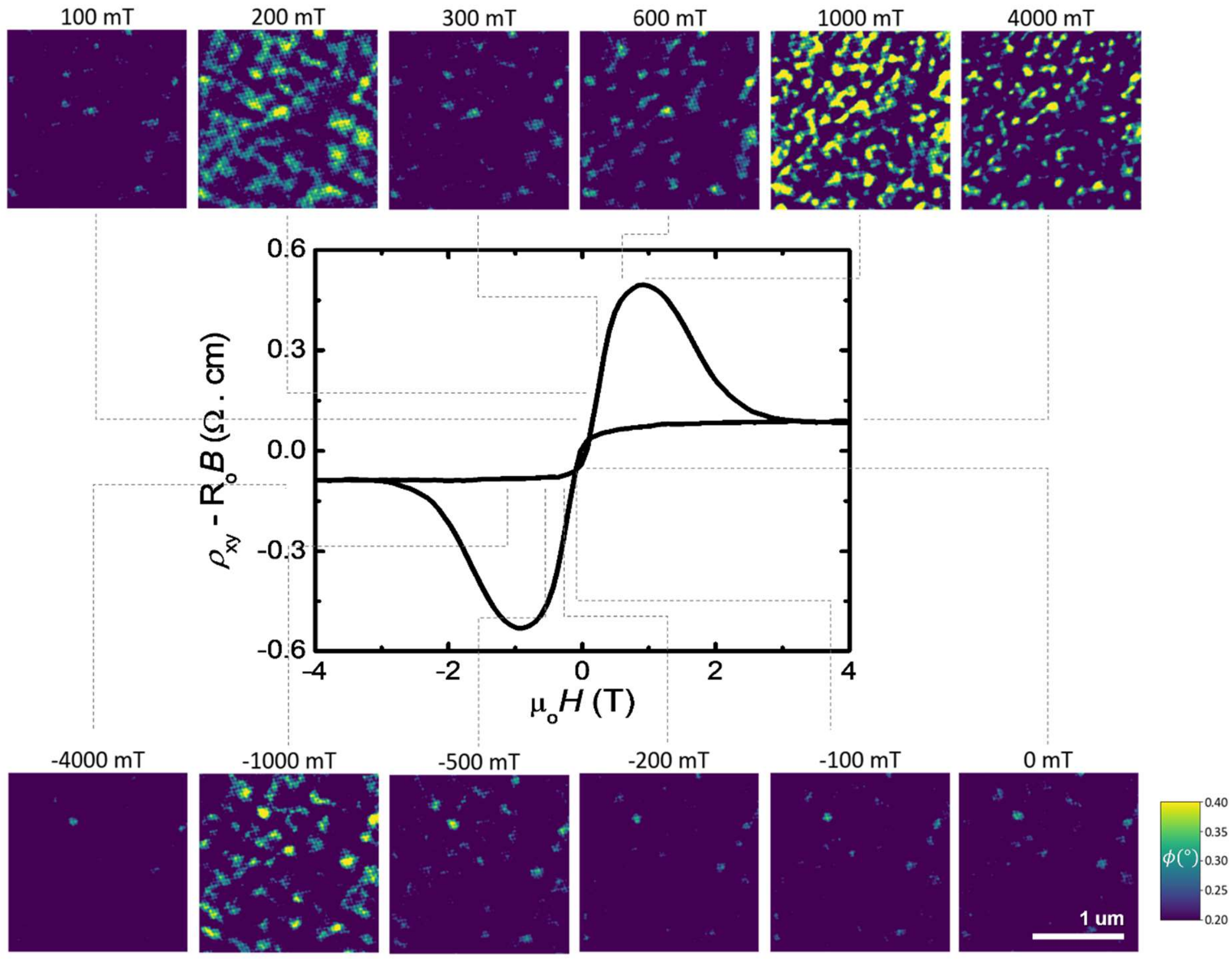

**Supporting Figure S2:** Magnetic Force Microscopy (MFM) $\Delta f$-imaging data on $mR_5I_{10}$. The magnetic field was swept from +4 Tesla to -4 Tesla, scanned at 20 K. The $\rho_{xy}(H_{\text{total}}, \theta = 0)$ at 20 K is inserted at the centre for comparison, where dotted lines indicate the corresponding images and Hall states.

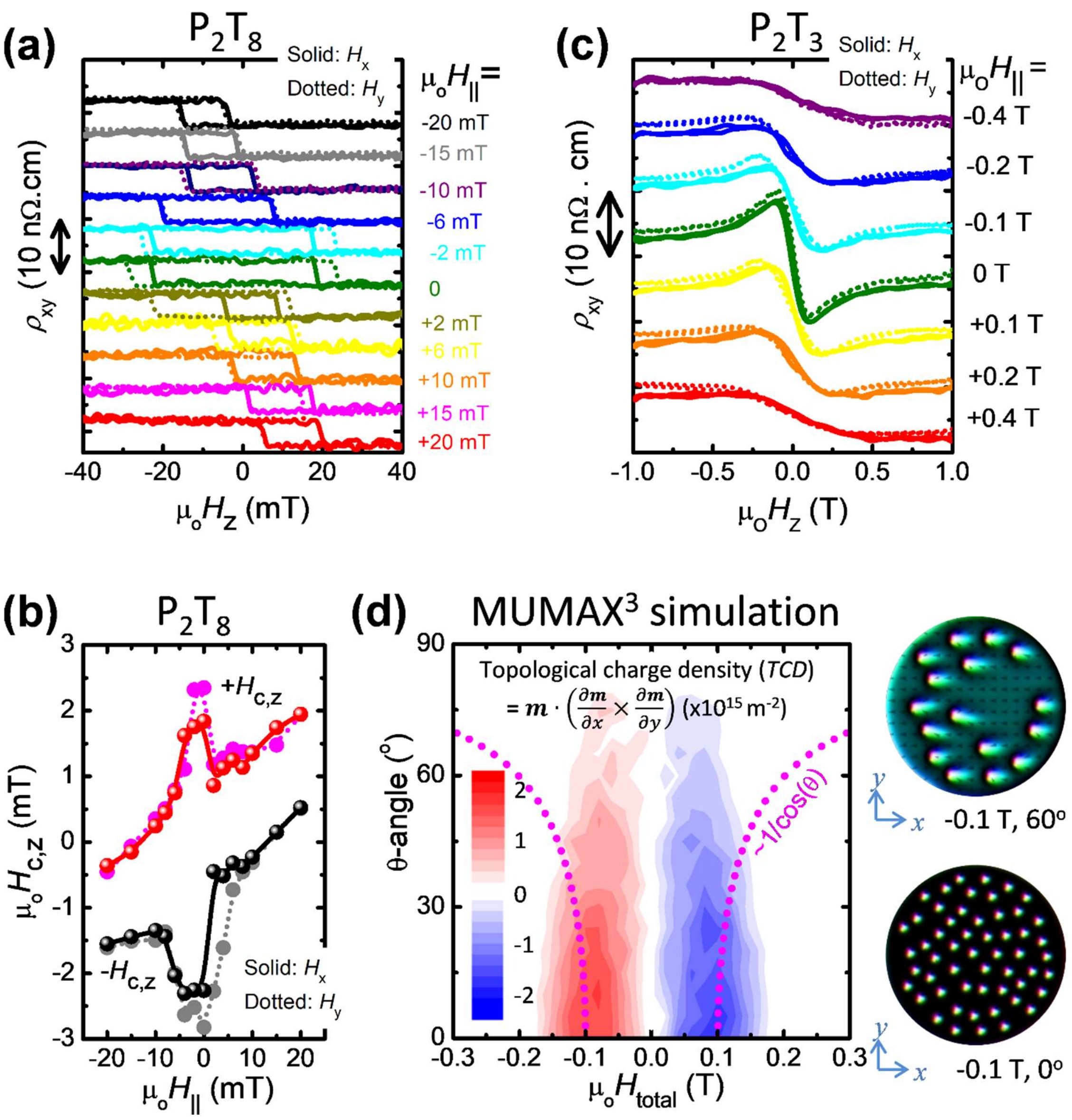

**Supporting Figure S3:** $\rho_{xy}(H_z, H_{x,y})$ data for **(a)** $P_2T_8$ and **(c)** $P_2T_3$, with comparison between $H_x$ (solid) and $H_y$ (dotted) in-plane fields. Loops are vertically shifted for clarity. **(b)** Summary of z-component $H_{peak}$ variations with $H_{x,y}$ in $P_2T_8$ from (a).. The charge current density $\boldsymbol{J}_{c,x}$ used was +7.5x10$^9$ A/m$^2$ for both $P_2T_8$ and $P_2T_3$. **(d)** *TCD*($H_{total}$,θ) mapping obtained by MUMAX$^3$ simulations done with parameters of $P_2T_3$, agreeing with its $\rho_{xy}$($H_{total}$,θ) data in Figure S2d. Right panels: snapshots of magnetic textures extracted from the ($H_{total}$,θ) states indicated.

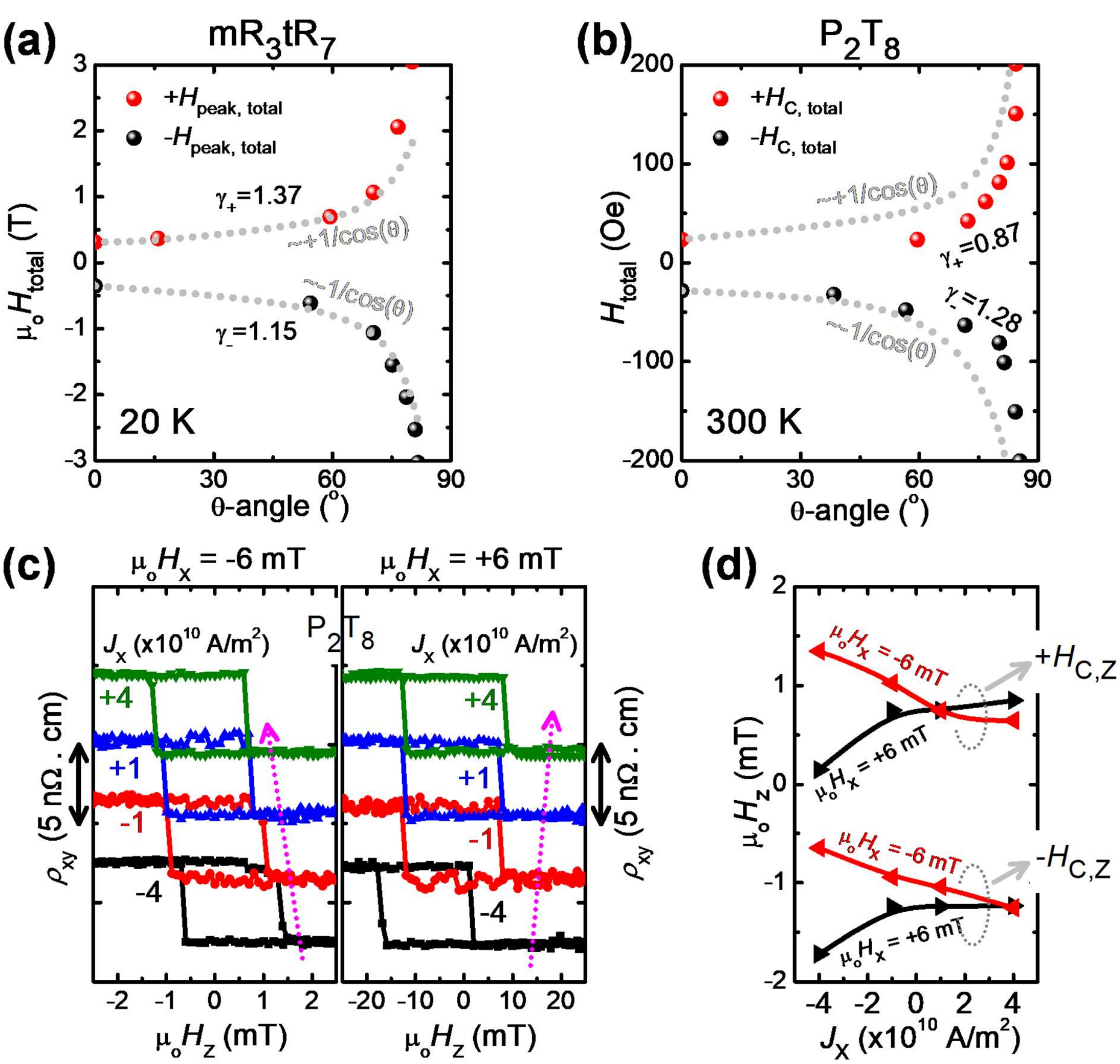

**Supporting Figure S4:** The resulted θ-angle dependent $H_{\text{peak,total}}$ or $H_{\text{C,total}}$ trends after transforming the $H_{\text{peak,z}}$ and $H_{\text{C,z}}$ data from the resolved field-component scheme in main text Figure 3 for **(a)** $mR_3tR_7$ and **(b)** $P_2T_8$. The γ-exponents for the positive and negative branches are labelled. In $P_2T_8$, no sign of reduction/shrinkage can be seen in $H_{\text{C,total}}$ with increasing θ, although reduction in $H_{\text{C,z}}$ with increasing $H_x$ is observed at small $H_x$. **(c)** $\rho_{xy}$ versus $H_z$ Hall-loops at constant $H_x$ = ±60 Oe but with varying $J_x$ from -4x10$^{10}$ to +4x10$^{10}$ A/m$^2$ showing clear shifts, with the z-component coercive fields ($H_{c,z}$) plotted in **(d)**. The shifting direction reverses upon reversing the sign of $H_x$ is a clear signature of SOT effect.

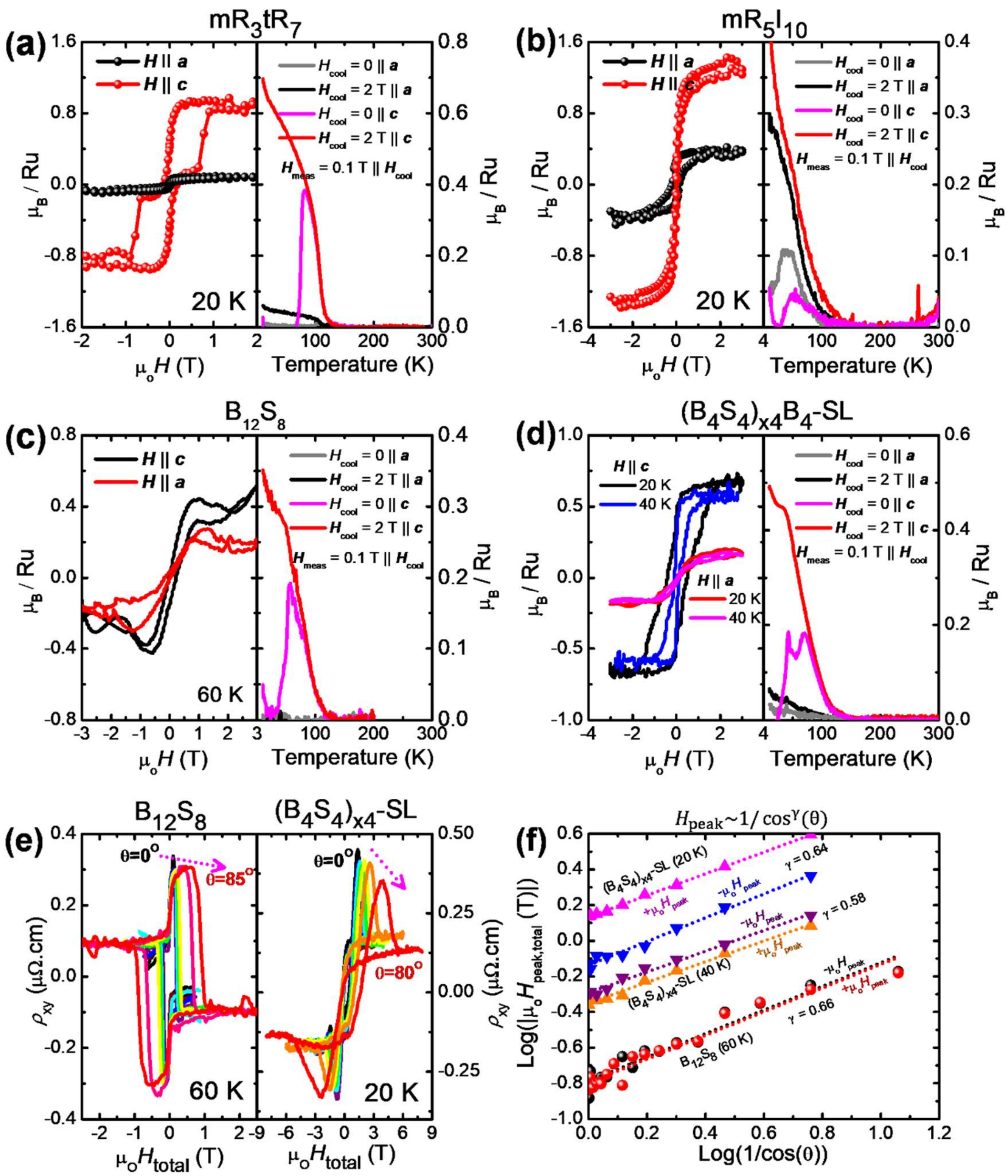

**Supporting Figure S5:** Magnetometry data for **(a)** $mR_3tR_7$ **(b)** $mR_5I_{10}$ **(c)** $B_{12}S_8$ and **(d)** $(B_4S_4)_{x4}$-SL respectively. In each heterostructure, the *M*-*H* curves for in-plane and out-of-plane fields are plotted on the left panels for calculations of $K_U$ at the temperature indicated, while *M*-*T* are plotted on the right panels for calculation of $A_{ex}$ from extracted $T_C$. **(e)** Hall-loops with increasing θ-angle in the total-field rotation scheme for $B_{12}S_8$ at 10 K (left) and $(B_4S_4)_{x4}$-SL at 20 K (right). **(f)** Log-log plots for γ extractions for various heterostructures after linear fittings (dotted lines).

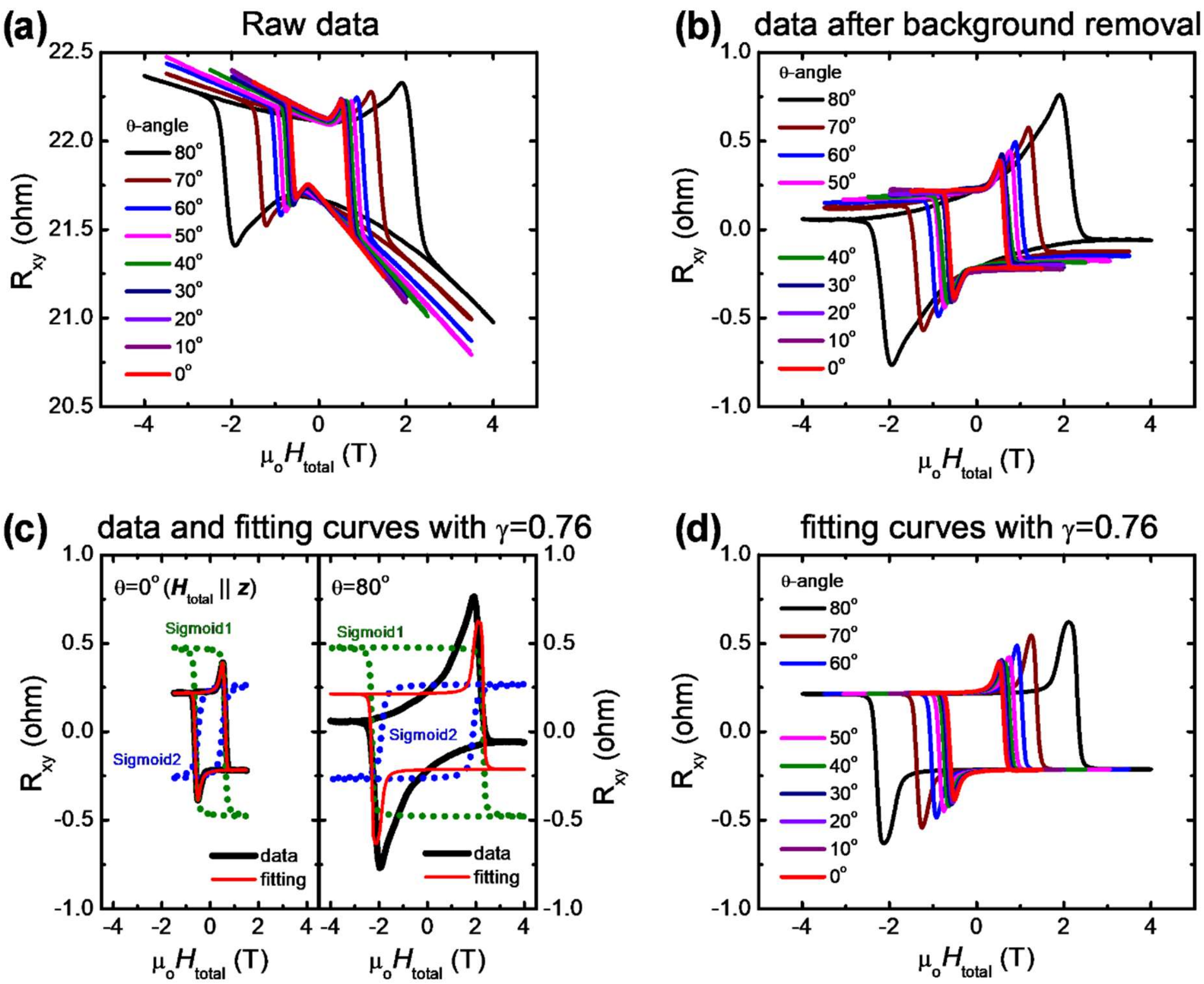

**Supporting Figure S6:** **(a)** raw data of Hall resistance from $mR_3tR_7$ (20 K) without any processing. **(b)** data of (a) after background removal. **(c)** Fitting of the Hall-humps data $mR_3tR_7$ by using the method of two overlapping sigmoidal KL-AHE loops for θ=0º (left panel) and 80º (right panel) respectively. The equation used is the same as shown in main text, i.e.: $\text{sigmoid}_{1,2} = A_{1,2} \coth\left\{B_{1,2}\left[\mu_o H \pm \frac{\mu_o H_{C1,2}}{(\cos\ )^{0.76}}\right]\right\} - \frac{A_{1,2}}{B_{1,2}\left[\mu_o H \pm \frac{\mu_o H_{C1,2}}{(\cos\theta)^{0.76}}\right]}$. **(d)** The fitting loops of all angles to be compared to (b) for similar trends. Note that the $\text{sigmoid}_{1,2}$ of all angles use the same fitting parameters as those of θ=0º, i.e.: $A_1$ =-0.48 ohm, $B_1 = 5.5\text{x}10^{-3}$, $\mu_o H_{C1}$ = 0.61 T, $A_2$ =+0.27 ohm, $B_2 = 1.65\text{x}10^{-3}$, $\mu_o H_{C2}$ = 0.5 T, and the common γ=0.76, only θ is varied. However, the fitting results at large θ are not good.

## SUPPORTING REFERENCES

[1] a) A. M. Glazer, *Acta Crystallographica Section B* **1972**, 28, 3384; b) A. M. Glazer, *Acta Crystallographica Section A* **1975**, 31, 756.

[2] T. C. van Thiel, J. Fowlie, C. Autieri, N. Manca, M. Šiškins, D. Afanasiev, S. Gariglio, A. D. Caviglia, *ACS Materials Letters* **2020**, 2, 389.

[3] a) Y. F. Nie, P. D. C. King, C. H. Kim, M. Uchida, H. I. Wei, B. D. Faeth, J. P. Ruf, J. P. C. Ruff, L. Xie, X. Pan, C. J. Fennie, D. G. Schlom, K. M. Shen, *Physical Review Letters* **2015**, 114, 016401; b) P. Schütz, D. Di Sante, L. Dudy, J. Gabel, M. Stübinger, M. Kamp, Y. Huang, M. Capone, M. A. Husanu, V. N. Strocov, G. Sangiovanni, M. Sing, R. Claessen, *Physical Review Letters* **2017**, 119, 256404.

[4] W. Lu, P. Yang, W. D. Song, G. M. Chow, J. S. Chen, *Physical Review B* **2013**, 88, 214115.

[5] A. K. H. K. Zhi Shiuh Lim, Zhou Zhou, Ganesh Ji Omar, Ping Yang, Robert Laskowski, Ariando Ariando, *arXiv:2204.06174* **2022**, DOI: arXiv:2204.06174.

[6] a) R. Laskowski, G. K. H. Madsen, P. Blaha, K. Schwarz, *Physical Review B* **2004**, 69, 140408; b) P. Blaha, K. Schwarz, F. Tran, R. Laskowski, G. K. H. Madsen, L. D. Marks, *The Journal of Chemical Physics* **2020**, 152, 074101.

[7] H. Yang, A. Thiaville, S. Rohart, A. Fert, M. Chshiev, *Phys. Rev. Lett.* **2015**, 115, 267210.

[8] L. M. Sandratskii, *Advances in Physics* **1998**, 47, 91.

[9] M. Heide, G. Bihlmayer, S. Blügel, *Physica B: Condensed Matter* **2009**, 404, 2678.

[10] C. Tang, P. Sellappan, Y. Liu, Y. Xu, J. E. Garay, J. Shi, *Phys. Rev. B* **2016**, 94, 140403.

[11] R. Pauthenet, **1958**, 29, 253.

[12] M. Raju, A. P. Petrović, A. Yagil, K. S. Denisov, N. K. Duong, B. Göbel, E. Şaşıoğlu, O. M. Auslaender, I. Mertig, I. V. Rozhansky, C. Panagopoulos, *Nat. Commun.* **2021**, 12, 2758.

[13] L. Caretta, E. Rosenberg, F. Büttner, T. Fakhrul, P. Gargiani, M. Valvidares, Z. Chen, P. Reddy, D. A. Muller, C. A. Ross, G. S. D. Beach, *Nature Communications* **2020**, 11, 1090.

[14] Z. S. Lim, C. Li, Z. Huang, X. Chi, J. Zhou, S. Zeng, G. J. Omar, Y. P. Feng, A. Rusydi, S. J. Pennycook, T. Venkatesan, A. Ariando, *Small* **2020**, 16, 2004683.

[15] L. Wang, Q. Feng, Y. Kim, R. Kim, K. H. Lee, S. D. Pollard, Y. J. Shin, H. Zhou, W. Peng, D. Lee, W. Meng, H. Yang, J. H. Han, M. Kim, Q. Lu, T. W. Noh, *Nature Materials* **2018**, 17, 1087.